\documentclass[10pt]{article}
\usepackage[utf8]{inputenc}
\usepackage{amsmath}
\usepackage{amssymb}
\usepackage{amsthm}
\usepackage{amsfonts}
\usepackage{graphicx}
\usepackage{todonotes}
\usepackage{natbib}
\usepackage{xcolor}
\usepackage{import}
\usepackage{setspace}
\usepackage{tikz}
\usetikzlibrary{arrows,positioning,shapes.geometric}
\presetkeys{todonotes}{size=\setstretch{1}\selectfont\tiny}{}
\date{\today}
\usepackage[margin=1in]{geometry}
\usepackage{float}
\usepackage{hyperref}

\DeclareMathOperator{\E}{\text{E}}

\newbox\keywdbox
\def\keywords{\global\setbox\keywdbox\vbox\bgroup\hsize\textwidth\small\leftskip0pc\rightskip\leftskip\noindent{\sc
Key words:\hskip1em}\ignorespaces}
\def\endkeywords{\egroup}

\newcommand{\bx}{\mathbf{X}}
\newcommand{\by}{\mathbf{y}}

\newcommand{\xtr}{X^\mathrm{train}}
\newcommand{\xte}{X^\mathrm{test}}

\newcommand{\xo}{X^{(1)}}
\newcommand{\xm}{X^{(m)}}
\newcommand{\xM}{X^{(M)}}
\newcommand{\bxo}{\mathbf{X}^{(1)}}
\newcommand{\bxm}{\mathbf{X}^{(m)}}
\newcommand{\bxM}{\mathbf{X}^{(M)}}

\newcommand{\bxmm}{\mathbf{X}^{(-m)}}
\newcommand{\xmm}{X^{(-m)}}

\newcommand{\xt}{X^{(2)}}

\DeclareMathOperator{\NB}{\mathrm{NB}}
\DeclareMathOperator{\bb}{\mathrm{BetaBinomial}}

\DeclareMathOperator{\Var}{\mathrm{Var}}

\DeclareMathOperator{\poi}{\mathrm{Poisson}}

\DeclareMathOperator{\eps}{\epsilon}
 
\newtheorem{algorithm}{Algorithm}
\newtheorem{theorem}{Theorem}

\newtheorem{example}{Example}

\title{Negative binomial count splitting \\
for single-cell RNA sequencing data}

\author{Anna Neufeld$^{1,2*}$, 
Joshua Popp$^{3}$, Lucy L. Gao$^{4}$, Alexis Battle$^{3,5}$, and Daniela Witten$^{1,6}$ \\
$^{1}$ Department of Statistics, University of Washington, Seattle, WA, USA \\
$^{2}$ Public Health Sciences Division, Fred Hutchinson Cancer Center, Seattle, WA, USA \\
$^{3}$ Department of Biomedical Engineering, Johns Hopkins University, Baltimore, Maryland, USA  \\
$^{4}$ Department of Statistics, University of British Columbia, Vancouver, BC, Canada \\
$^{5}$ Department of Computer Science, Department of Genetic Medicine, and Malone Center \\ for Engineering in Healthcare, Johns Hopkins University, Baltimore, Maryland, USA \\
$^{6}$ Department of Biostatistics, University of Washington, Seattle, WA, USA  \\
\textit{*email:} aneufeld@fredhutch.org
}

\begin{document}
\maketitle

\begin{abstract}
The analysis of single-cell RNA sequencing (scRNA-seq) data often involves fitting a latent variable model to learn a low-dimensional representation for the cells. Validating such a model poses a major challenge. If we could sequence the same set of cells twice, we could use one dataset to fit a latent variable model and the other to validate it. In reality, we cannot sequence the same set of cells twice. \emph{Poisson count splitting} was recently proposed as a way to work backwards from a single observed Poisson data matrix to obtain independent Poisson training and test matrices that could have arisen from two independent sequencing experiments conducted on the same set of cells. However, the Poisson count splitting approach requires that the original data are exactly Poisson distributed: in the presence of any overdispersion, the resulting training and test datasets are not independent.  In this paper, we introduce \emph{negative binomial count splitting}, which extends Poisson count splitting to the more flexible negative binomial setting. Given an $n \times p$ dataset from a negative binomial distribution, we use Dirichlet-multinomial sampling to create two or more independent $n \times p$ negative binomial datasets. We show that this procedure outperforms Poisson count splitting in simulation, and apply it to validate clusters of kidney cells from a human fetal cell atlas. 
\end{abstract}


\section{Introduction}
\label{section_intro_nb}

A single-cell RNA sequencing (scRNA-seq) dataset involving $n$ cells and $p$ genes can be written as a matrix $X \in \mathbb{Z}_{\geq 0}^{n \times p}$, where entry $X_{ij}$ is the number of unique molecular identifiers from the $i$th cell that map to the $j$th gene. It is common to model $X$ as a realization from a random variable $\bold{X}$, and assume that
\begin{equation}
\label{eq_meanmodel}
\E\left[ \bx_{ij} \right] = \gamma_i \Lambda_{ij}, \text{ with }g(\Lambda) = L \beta^\top \text{ for } L \in \mathbb{R}^{n \times K^*}, \beta \in \mathbb{R}^{p \times K^*},
\end{equation}
for some link function $g(\cdot)$ \citep{sarkar2021separating}. In \eqref{eq_meanmodel}, $\gamma=(\gamma_1,\ldots,\gamma_n)^T$ stores cell-specific \emph{size factors}, which reflect technical variation in sequencing depth between cells. The matrix $\Lambda$ represents the biological variation of interest, which, after applying a link function $g(\cdot)$, has rank $K^*$ for some $K^* \leq \min(n,p)$. 

Fitting the model \eqref{eq_meanmodel}--that is, obtaining estimates $\hat{L}(X)$ and $\hat{\beta}(X)$ of $L$ and $\beta$--may be of interest for a number of reasons. For example, we may wish to denoise the data by replacing the size-factor normalized dataset $\mathrm{diag}(\gamma)^{-1} X$ with its low-rank estimate $g^{-1}\left(\hat{L}(X)\hat{\beta}(X)^T \right)$ \citep{eraslan2019single, townes2019feature}, or we may wish to interpret $\hat{L}(X)$ as a measure of an unobserved aspect of cell state, e.g. cell type or position along a developmental trajectory \citep{aizarani2019human, van2020trajectory}. After fitting the model \eqref{eq_meanmodel}, we typically want to perform some type of model validation or inference. 

\begin{example}
\label{ex:denoise}
We want to assess the quality of our low-rank approximation $\mathrm{diag}(\gamma)^{-1} X \approx g^{-1}\left( \hat{L}(X)\hat{\beta}(X)^T\right)$ \citep{sarkar2021separating, batson2019molecular}. 
\end{example}

\begin{example}
\label{ex:diffExp}
We want to identify genes that are associated with $\hat{L}(X)$ \citep{aizarani2019human, van2020trajectory, zhang2019valid}. 
\end{example}

\begin{example}
\label{ex:stable}
We want to know whether we would obtain a similar estimate  $\hat{L}(X)$ on an independent realization of $\bold{X}$ drawn from the same distribution \citep{cao2020human, lange2004stability, ullmann2022validation}. We refer to this as ``reproducibility."
\end{example}

In Example~\ref{ex:denoise}, \emph{because we estimated $L$ and $\beta$ on the data $X$, we cannot re-use $X$ to assess model fit} \citep{hastie2009elements}. In Example~\ref{ex:diffExp}, \emph{because we estimated $L$ on the data $X$, we cannot re-use $X$ to test for association} \citep{gao2022selective}. In Example~\ref{ex:stable}, \emph{we only have access to one dataset}, so it is unclear how to proceed. While very specialized approaches are available to overcome these challenges in specific instantiations of Example~\ref{ex:denoise} \citep{fu2020estimating, grabski2022significance}, Example~\ref{ex:diffExp} \citep{gao2022selective, chen2022selective, zhang2019valid, chung2015statistical}, and Example~\ref{ex:stable} \citep{tibshirani2005cluster, lange2004stability}, in this paper we will provide a much more flexible framework for model validation or inference after fitting \eqref{eq_meanmodel}, which will be applicable to all three examples. 

\begin{figure}
\includegraphics[width=\textwidth]{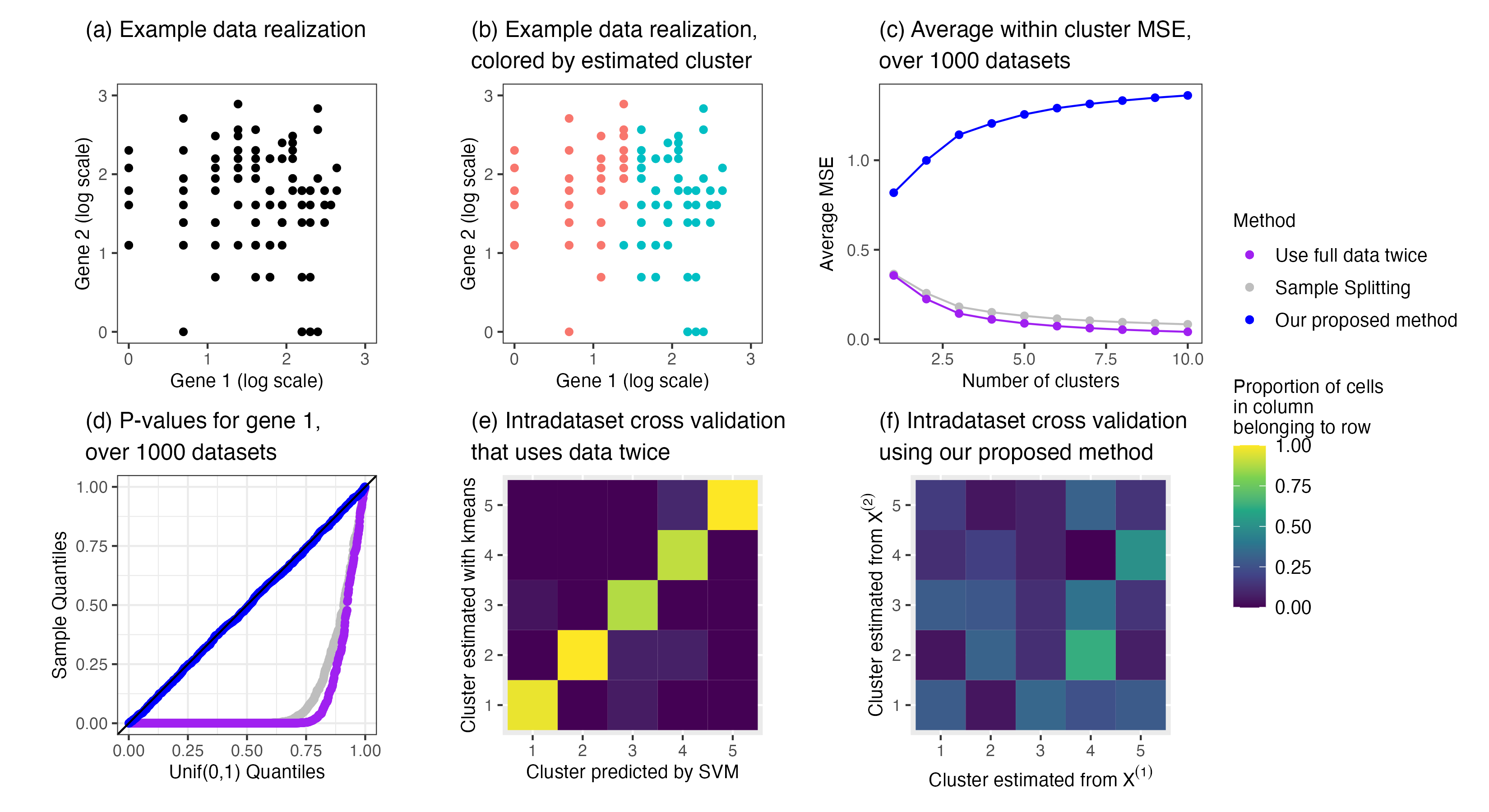}
\caption{
\emph{(a)} Data $X \in \mathbb{Z}_{\geq 0}^{100 \times 2}$, where each entry $X_{ij}$ is drawn independently from a negative binomial distribution with mean $5$ and variance $10$. \emph{(b)} The same data $X$, colored by the clusters estimated when $k$-means with $K=2$ is applied to $\log(X+1)$. 
\emph{(c)}  The within-cluster MSE computed after fitting k-means with $K = 1,2,\ldots,10$, averaged over 1000 datasets. We consider three approaches: we cluster using all of the data and then compute the MSE using all of the data (purple), we cluster using 50 observations and compute the MSE on the other 50 (gray), and we apply our proposed negative binomial count splitting method (blue). Details are given in Section~\ref{subsec_estK}.
\emph{(d)} Uniform QQ-plot of negative binomial GLM p-values for testing for differential expression of gene 1 across the estimated clusters for $1,000$ realizations of $X$. We consider three approaches: we cluster and fit the GLM using all of the data (purple), we cluster using 50 observations and fit the GLM on the other 50 (gray), and we apply our proposed negative binomial count splitting method (blue). Details are given in Section~\ref{subsec_difExp}.
\emph{(e)} Normalized confusion matrix resulting from the intradataset cross-validation procedure of \protect\cite{cao2020human} (see Algorithm~\ref{alg_cao_intra} in Section~\ref{section_realData_nb}) that uses the same data for both clustering and cross-validation. 
\emph{(f)} Normalized confusion matrix resulting from our modified version of intradataset cross-validation (see Algorithm~\ref{alg_cs_intra} in Section~\ref{section_realData_nb}).}
\label{fig_intro}
\end{figure}

To illustrate the problem, we generate a toy data matrix $X \in \mathbb{Z}_{\geq 0}^{100 \times 2}$ with elements drawn independently from a negative binomial distribution with mean $5$ and variance $10$. This is a special case of \eqref{eq_meanmodel} with $\gamma_1 = \ldots = \gamma_n = 1$, $K^*=1$, $L = \bold{1}_{100}$, and $\beta = [5, 5]^T$. The data are shown in Figure~\ref{fig_intro}(a). To illustrate Example~\ref{ex:denoise}, we apply
$k$-means clustering to the data to estimate $K$ clusters for a range of values of $K$. 
Though there is one true cluster in this example (all 100 cells are homogenous), the mean squared error (MSE; defined in \eqref{eq_mse} in Section 4) computed on the same data used for clustering is monotone decreasing in $K$ (Figure~\ref{fig_intro}(c)), incorrectly suggesting that a larger value of $K$ always leads to a better fit. To illustrate Example~\ref{ex:diffExp}, we fit a negative binomial generalized linear model (GLM) to test whether the expected expression of the first gene is associated with the cluster labels when we estimate $K=2$ clusters. When we perform this test on the same data used for clustering, we obtain p-values that are much smaller than the $\mathrm{Unif}(0,1)$ distribution (Figure~\ref{fig_intro}(d)), and thus do not control the Type 1 error rate (recall that no true clusters are present, and thus there is no true association between the first gene and the estimated clusters). 

The solution here might seem obvious: to split our 100 cells into a training set, used to fit \eqref{eq_meanmodel}, and a test set, used for model validation. Unfortunately, this \emph{sample splitting} approach does not work. The issue is that fitting \eqref{eq_meanmodel} using the cells in the training set yields latent variable coordinates for cells in the training set only. To use the test set for validation or inference, we must obtain latent variable coordinates of the cells in the test set. This step involves using the test set data itself, which invalidates downstream evaluation or inference. 
In our toy example in Figure~\ref{fig_intro}, we apply $k$-means clustering to the cells in the training set, and then assign cluster labels to the cells in the test set using 3-nearest neighbor classification. We see in Figure 1(c) that, over 1000 simulated datasets, the within-cluster MSE computed on the test set (see Appendix~\ref{appendix_naive_sampsplit} for details) decreases monotonically with the number of clusters, because we used the test set both to compute latent variable coordinates for the cells in the test set and to compute the within-cluster MSE. Similarly, Figure~\ref{fig_intro}(d) shows that, over 1,000 simulated datasets, the p-values from a negative binomial GLM that regresses the first gene from the test set onto the test set cluster assignments do not control the Type 1 error rate. We refer the reader to \cite{owen2009bi} for more discussion of the inadequacy of sample splitting in the setting of Example~\ref{ex:denoise}, and \cite{gao2022selective}, \cite{chen2022selective}, and \cite{neufeld2022inference} for a related discussion in the setting of Example~\ref{ex:diffExp}.

In the setting of Example~\ref{ex:stable}, \cite{cao2020human} implement a procedure that they call ``intradataset cross-validation" (see Algorithm~\ref{alg_cao_intra}). Inspired by the general framework of \cite{abdelaal2019comparison},  their procedure involves estimating clusters using all of the data and then performing 5-fold cross-validation to assess the accuracy of a classifier fit to predict these clusters. Low cross-validation error is treated as evidence of cluster reproducibility, because it means that a given cell's cluster assignment can be reproduced by a classifier, even when that cell itself was not used to train the classifier. Figure~\ref{fig_intro}(e) shows a confusion matrix comparing the cell types estimated via clustering (with $K=5$) to those predicted using cross-validation (with five folds and a support vector machine classifier) for the cells in the toy dataset from Figure~\ref{fig_intro}(a). Despite the fact that all cells are homogenous in this dataset (and thus the estimated clusters are driven by random noise), 95\% of the cells fall on the diagonal of the confusion matrix, falsely suggesting reproducibility of the clusters. The issue is that, since all of the data from all of the cells was used for clustering, any downstream model evaluation is compromised, even if the downstream task makes use of cross-validation. 

Now, suppose that we were able to sequence the same set of cells twice to obtain two independent datasets $\xtr \in \mathbb{Z}_{\geq 0}^{n \times p}$ and $\xte \in \mathbb{Z}_{\geq 0}^{n \times p}$ generated from $\bold{X}$ in \eqref{eq_meanmodel} (with the same true underlying $L$ and $\beta$ matrices). We could estimate $L$ and/or $\beta$ using only $\xtr$, and could then validate the results or conduct inference using $\xte$. Thus, the challenges associated with Examples 1 and 2 displayed in Figure 1(c) and 1(d) would be entirely avoided. Similarly, we could estimate one set of clusters on  $\xtr$ and another set of clusters on  $\xte$ and compare the two clusterings using a metric such as the adjusted Rand Index \citep{hubert1985comparing}, entirely avoiding the challenge of Example 3. 

In practice, we cannot sequence the same set of cells twice. Instead, we propose using our single dataset $X$ to reverse engineer two datasets $\xtr$ and $\xte$ that function like independent sequencing experiments performed on the same sets of cells. Our proposal is an extension of the ideas of \cite{batson2019molecular}, \cite{sarkar2021separating}, and \cite{neufeld2022inference}, who perform this reverse engineering under the assumption that $X_{ij} \overset{\mathrm{ind.}}{\sim} \mathrm{Poisson}(\Lambda_{ij})$. However, scRNA-seq data are typically overdispersed relative to the Poisson or the binomial distribution, and so the \emph{Poisson count splitting} procedure developed in these earlier papers will fail to produce independent training and test sets. 

In recent work, \cite{neufeld2023data} and \cite{dharamshi2023generalized} developed \emph{data thinning}, a vast generalization of Poisson count splitting that enables us to split a random variable drawn from a number of well-known distributional families into two or more independent components. In this paper, we focus on the special case of data thinning for negative binomial random variables, which allows us to obtain independent matrices $\xtr$ and $\xte$ under the assumption that the elements of $X$ are independent draws from negative binomial distributions. We refer to this procedure as \emph{negative binomial count splitting}. Critically, $\xtr$ and $\xte$ are drawn from the same distribution as $X$, up to a parameter scaling. 

Figure~\ref{fig_intro}(c) and Figure~\ref{fig_intro}(d) show that, in our toy example, negative binomial count splitting correctly determines that $K^*=1$ and controls the Type 1 error rate. Figure~\ref{fig_intro}(f) shows that a modified version of ``intradataset cross-validation" that makes use of negative binomial count splitting yields a confusion matrix that accurately reflects the absence of signal. 

Negative binomial count splitting requires a negative binomial assumption to ensure independence between the training and test sets. However, once the data has been split, we are free to use any latent variable estimation method or inferential technique. 

In Section~\ref{section_background_nb}, we review the Poisson count splitting procedure of \cite{neufeld2022inference} and its generalization to data thinning \citep{neufeld2023data, dharamshi2023generalized}. In Section~\ref{section_gcs}, we introduce negative binomial count splitting and provide some theoretical results. In Section~\ref{section_simulation_nb}, we apply negative binomial count splitting to Example~\ref{ex:denoise} and Example~\ref{ex:diffExp} on simulated data. In Section~\ref{section_realData_nb}, we revisit the intradataset cross-validation procedure of \cite{cao2020human}, and assess the reproducibility (Example~\ref{ex:stable}) of cell types and subtypes from their human cell atlas using negative binomial count splitting. We close with a brief discussion in Section~\ref{sec_disc}.

\section{Background}
\label{section_background_nb}

\subsection{A review of data thinning}
\label{subsec_csreview}

Our goal is to decompose an scRNA-seq dataset $X \in \mathbb{Z}_{\geq 0}^{n \times p}$ into independent datasets  $\xtr \in \mathbb{Z}_{\geq 0}^{n \times p}$ and $\xte \in \mathbb{Z}_{\geq 0}^{n \times p}$ drawn from the same model as $X$, up to a parameter scaling. 
We first review \emph{Poisson count splitting}, which accomplishes this goal if $\bold{X}_{ij} \overset{\mathrm{ind.}}{\sim} \mathrm{Poisson}(\Lambda_{ij})$. 

\begin{algorithm}[Poisson count splitting]
\label{alg_cs}
Let $X \in \mathbb{Z}_{\geq 0}^{n \times p}$. For a chosen $M \in \mathbb{Z}^{+}$ and $\epsilon_1,\ldots,\epsilon_M \in (0,1)$ such that $\sum_{m=1}^M \epsilon_m = 1$, draw $\left(\bold{X}_{ij}^{(1)},\bold{X}_{ij}^{(2)}, \ldots, \bold{X}_{ij}^{(M)}\right) \mid \bold{X}_{ij} = X_{ij} \sim \mathrm{Multinomial}\left(X_{ij}, \epsilon_1,\ldots, \epsilon_M\right)$.
\end{algorithm}

\begin{theorem}
\label{theorem_poithin}	
Let $X \in \mathbb{Z}_{\geq 0}^{n \times p}$ be a dataset with entries $X_{ij}$ drawn from  $\bold{X}_{ij} \overset{\mathrm{ind.}}{\sim} \poi(\mu_{ij})$. If we apply Algorithm~\ref{alg_cs} to each element $X_{ij}$ of this data, then
(1) $\bxm_{ij} \overset{\mathrm{ind.}}{\sim} \mathrm{Poisson}(\epsilon_m \mu_{ij})$ for $i=1,\ldots,n$, $j=1,\ldots,p$, and (2) the folds $\bxo, \ldots, \bxM$ are mutually independent.
\label{theorem_poithin}
\end{theorem}

Theorem~\ref{theorem_poithin} follows from the well-known binomial thinning property of the Poisson distribution (see \citealt{durrett2019probability}, Section 3.7.2). For any fold $m \in \{ 1,\ldots,M\}$, we define $\bxmm = \bx - \bxm$. The mutual independence between folds given in Theorem~\ref{theorem_poithin} ensures that, for all $m \in \{ 1,\ldots,M\}$, $\bxm$ is independent of $\bxmm$. Thus, by treating $\bxmm$ as a training set and $\bxm$ as a test set, we arrive at an alternative to cross-validation that uses Poisson count splitting, rather than sample splitting, to create training and test sets. 

\cite{sarkar2021separating} and \cite{neufeld2022inference} apply Poisson count splitting to overcome the challenges arising in Examples 1 and 2 of Section~\ref{section_intro_nb}, under the assumption that the scRNA-seq data follows a Poisson distribution. Unfortunately, the 
Poisson assumption in Theorem~\ref{theorem_poithin} is necessary to achieve independence between the folds $\bxo, \ldots, \bxM$.  If the data instead follow a negative binomial distribution, then $\bxo, \ldots, \bxM$ are correlated, and Poisson count splitting will fail to provide a valid approach for model evaluation or inference. 

\cite{neufeld2023data} and \cite{dharamshi2023generalized} expand upon Poisson count splitting to describe a general recipe for decomposing a single random variable $\bold{Y} \sim F_{\theta}$, for some distribution $F_\theta$ indexed by an unknown parameter $\theta$, into independent pieces $\bold{Y}^{(m)} \overset{\mathrm{ind.}}{\sim} Q^{(m)}_{\theta}$, where $Q^{(m)}_\theta$ is a (possibly different) distribution indexed by the same parameter $\theta$. They refer to this general framework as \emph{data thinning}. In particular, the framework developed by \cite{neufeld2023data} enables us to decompose a negative binomial random variable into $M \geq 2$ independent negative binomial random variables, without knowledge of the mean parameter. While the property of the Poisson distribution that allows for Poisson count splitting is well-known, its negative binomial counterpart is less well-known. In fact, prior to \cite{neufeld2023data}, thinning the negative binomial distribution appears to be unexplored outside of the time series literature of the 1990s \citep{joe1996time}. 


\subsection{Negative binomial models for scRNA-seq data}
\label{subsec_nb}

Throughout this paper, we let $\NB(\mu, b)$ denote the negative binomial distribution with mean $\mu$ and variance $\mu + \frac{\mu^2}{b}$, for $\mu > 0$ and $b > 0$. This parameterization is commonly used when modeling scRNA-seq data. Note that if $\by \mid \tau \sim \mathrm{Poisson}(\mu \tau)$ and $\tau \sim \mathrm{Gamma}(b,b)$, then $\by \sim \NB(\mu, b)$.  Because the variance $\mu + \frac{\mu^2}{b}$ is always strictly larger than the mean $\mu$, the negative binomial model is overdispersed relative to the Poisson model. This motivates its use in the analysis of RNA sequencing data, where the data are non-negative integers with excess variance relative to the Poisson distribution \citep{choudhary2022comparison, hafemeister2019normalization, sarkar2021separating, lopez2018deep}. We refer to the parameter $b$ as the overdispersion parameter. As $b \rightarrow \infty$, the negative binomial distribution approaches the Poisson distribution. In the scRNA-seq literature, it is common to assume that each gene, but not each cell, has its own overdispersion parameter \citep{hafemeister2019normalization, love2014moderated, lopez2018deep}. Thus, in what follows, we will assume that $\bold{X}_{ij} \overset{\mathrm{ind.}}{\sim} \mathrm{NB}\left( \mu_{ij}, b_j\right)$.

\section{Negative binomial count splitting}
\label{section_gcs}

In this section, we consolidate some theoretical results from \cite{neufeld2022inference} and \cite{neufeld2023data} to facilitate their immediate application to scRNA-seq data. Results in this section are proven in Appendix~\ref{appendix_proofs}.
 
\subsection{Algorithm and main result}

We now introduce negative binomial count splitting and state the key result. 

\begin{algorithm}[Negative binomial count splitting]
\label{alg_gcs}
Let $X \in \mathbb{Z}_{\geq 0}^{n \times p}$. For a chosen $M \in \mathbb{Z}^{+}$, $b_j' \geq 0$ for $j=1,\ldots,p$, and $\epsilon_1,\ldots,\epsilon_M \in (0,1)$ such that $\sum_{m=1}^M \epsilon_m = 1$, draw \\ $\left(\bx_{ij}^{(1)}, \ldots, \bx_{ij}^{(M)}\right) \mid \bx_{ij}=X_{ij} \sim \mathrm{DirichletMultinomial}\left(X_{ij}, \epsilon_1 b_j', \ldots, \epsilon_M b_j' \right)$. 
\end{algorithm}

The marginals of a Dirichlet-multinomial distribution are beta-binomial, i.e. $\bxm_{ij} \mid \bx_{ij} = X_{ij} \sim \text{BetaBinomial}(X_{ij}, \epsilon_m b_j', (1-\epsilon_m) b_j')$. While binomial thinning has appeared in numerous papers as a way to construct training and test sets from count-valued data \citep{neufeld2022inference, sarkar2021separating, leiner2021data}, to our knowledge beta-binomial thinning has only been used for this purpose by \cite{neufeld2023data}. The beta-binomial thinning operator has appeared in the time series literature as far back as \cite{mckenzie1986autoregressive} for constructing autoregressive processes with negative binomial marginals. The following result, which appeared for $M=2$ in the context of autoregressive processes in \cite{joe1996time}, tells us what happens when $\bx_{ij} \sim \mathrm{NB}\left(\mu_{ij}, b_j\right)$ and we apply Algorithm~\ref{alg_gcs} with $b_j'=b_j$.

\begin{theorem}
\label{theorem_nbthin}	
Let $X \in \mathbb{Z}_{\geq 0}^{n \times p}$ be a dataset such that the entries $X_{ij}$ are realizations of $\bx_{ij} \overset{\mathrm{ind.}}{\sim} \NB\left(\mu_{ij}, b_j \right)$. If we apply Algorithm~\ref{alg_gcs} to each element $X_{ij}$ of this data using $b_j'=b_j$,
then (1) $\bxm_{ij} \overset{\mathrm{ind.}}{\sim} \NB(\epsilon_m \mu_{ij}, \epsilon_m b_{j})$ for $i=1,\ldots, n$, $j=1,\ldots, p$, and (2) the folds $\xo, \ldots, \xM$ are mutually independent.
\end{theorem}

Drawing $(\bold{X}_{ij}^{(1)}, \ldots, \bold{X}_{ij}^{(M)}) \mid \bold{X}_{ij}=X_{ij}$ from a Dirichlet-multinomial distribution, as in Algorithm~\ref{alg_gcs}, is the same as first drawing $\left( \boldsymbol{\epsilon}_1, \ldots, \boldsymbol{\epsilon}_M \right)$ from a $\mathrm{Dirichlet}(\epsilon_1 b_j', \ldots, \epsilon_m b_j')$ distribution and then letting $(\bold{X}_{ij}^{(1)}, \ldots, \bold{X}_{ij}^{(M)})  \mid \bold{X}_{ij}=X_{ij} \sim \mathrm{Multinomial}(X_{ij}, \boldsymbol{\epsilon}_1, \ldots, \boldsymbol{\epsilon}_M)$. Thus, to accommodate the overdispersion in $\bold{X}_{ij}$ relative to the Poisson distribution, we add additional randomness to the sampling process by making the parameters $(\epsilon_1,\ldots, \epsilon_M)$ from Algorithm~\ref{alg_cs} random variables. 

Theorem~\ref{theorem_nbthin} implies that $\bxm$ is independent of $\bxmm : = \bx - \bxm$ for $m=1,\ldots,M$. To address Examples~\ref{ex:denoise}, \ref{ex:diffExp}, and \ref{ex:stable} from Section~\ref{section_intro_nb}, we will use $\bxmm$ as a training set and $\bxm$ as a test set. The bottom line is that if we believe that $\bold{X}_{ij} \overset{\mathrm{ind.}}{\sim} \NB\left(\mu_{ij}, b_j\right)$ and we know the true values $b_j$, then a direct extension of Poisson count splitting is available.

\subsection{The role of the parameter $b_j'$}

Theorem~\ref{theorem_nbthin} requires that we apply Algorithm~\ref{alg_gcs} with the correct value of the overdispersion parameter; i.e. that $\bx_{ij} \sim \mathrm{NB}(\mu_{ij}, b_j)$ and we choose $b_j' = b_j$. In this section, we consider what happens when $b_j' \neq b_j$.

When $b'_j = \infty$, drawing $\left(\bx_{ij}^{(1)}, \ldots, \bx_{ij}^{(M)}\right) \mid \bx_{ij}=X_{ij} \sim \mathrm{DirichletMultinomial}\left(X_{ij}, \epsilon_1 b_j', \ldots, \epsilon_M b_j' \right)$ is equivalent to  drawing $\left(\bx_{ij}^{(1)},\ldots, \bx_{ij}^{(M)}\right) \mid \bx_{ij}=X_{ij} \sim \mathrm{Multinomial}\left(X_{ij}, {\epsilon}_1, \ldots, {\epsilon}_M \right)$, and so Algorithm~\ref{alg_gcs} reduces to Algorithm~\ref{alg_cs}. 

\begin{theorem}[\cite{neufeld2022inference}]
\label{theorem_nb_binom_thin}	
If $\bx_{ij} \overset{\mathrm{ind.}}{\sim} \NB\left(\mu_{ij}, b_j\right)$ and we apply Algorithm~\ref{alg_gcs} with $b_j' = \infty$ to $\bx_{ij}$, then (1) $\bxm_{ij} \overset{\mathrm{ind.}}{\sim} \NB(\epsilon_m \mu_{ij},  b_j)$ and (2) $\mathrm{Cor}(\bxm_{ij}, \bxmm_{ij}) = \frac{\sqrt{\epsilon_m(1-\epsilon_m)}}{\sqrt{\frac{b_j^2}{\mu_{ij}^2} + \frac{b_j}{\mu_{ij}} + \epsilon_m(1-\epsilon_m)}}$, where $\bx^{(-m)} := \bx - \bx^{(m)}$. 
\end{theorem}

Theorem~\ref{theorem_nb_binom_thin} says that while applying Poisson count splitting (or negative binomial count splitting with $b_j'=\infty$) on data from a negative binomial distribution yields training and test sets that follow the same model as the full data up to a parameter scaling, these datasets are positively correlated. The positive correlation increases as the true value of $b_j$ decreases, and decreases to $0$ as $b_j \rightarrow \infty$. Moreover, we see from Theorem~\ref{theorem_nb_binom_thin} that the  overdispersion parameters (and thus the variances) of $\bxm_{ij}$ are too small relative to Theorem~\ref{theorem_nbthin}. Thus, by failing to put enough noise into our sampling process, applying Poisson count splitting to negative binomial data results in training and test sets that are not as noisy as they should be, leading to positive correlation between them. 

We now consider the more general case of finite $b_j'$. The following result is included (under a different parameterization) in \cite{neufeld2023data}. 

\begin{theorem}
\label{theorem_generalcase}
If $\bx_{ij} \sim \NB(\mu_{ij}, b_j)$ and we apply Algorithm~\ref{alg_gcs} with parameter $b_j'$:
\begin{enumerate}
\item $\E[\bxm_{ij}] = \epsilon_m \mu_{ij}$,
\item $\Var(\bxm_{ij}) = \epsilon_m \Var(\bx_{ij}) + \epsilon_m (1-\epsilon_m) \frac{\mu_{ij}^2}{b_j} \left( 
\frac{b_j+1}{b_j'+1} - 1\right)
$,
\item $\mathrm{Cov}(\bxm_{ij}, \bxmm_{ij}) = \epsilon_m (1-\epsilon_m) \frac{\mu_{ij}^2}{b_j} \left( 
1- \frac{b_j+1}{b_j'+1} \right)$, where $\bx^{(-m)} := \bx - \bx^{(m)}$. 
\end{enumerate}
\end{theorem}

\begin{figure}
\centering
\includegraphics[width=0.5\textwidth]{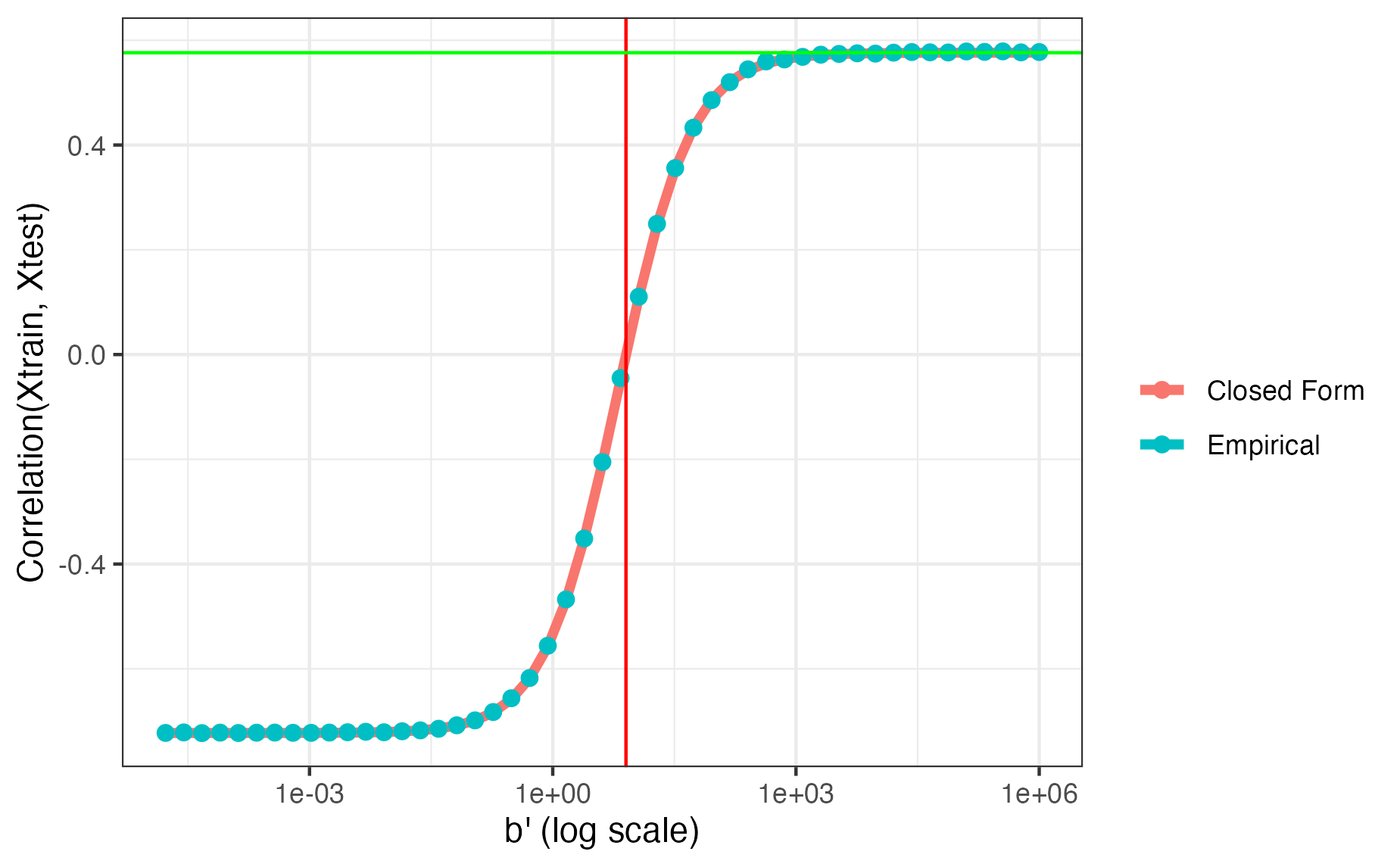}	
\caption{We generate $100,000$ independent realizations of $\bx_{11} \sim \mathrm{NB}(25,8)$. Then, for 50 values of $b_1'$ ranging from $10^{-6}$ to $10^{6}$, we split each of these realizations into $\bx^{(1)}_{11}$ and $\bx^{(2)}_{11}$ by applying Algorithm~\ref{alg_gcs} with $b_1', M = 2, \epsilon_1 = 0.3, \epsilon_2 = 0.7$. We display the sample correlation between the $100,000$ realizations of $\bx_{11}^{(1)}$ and $\bx_{11}^{(2)}$, as a function of $b'_1$. The pink line shows the theoretical values computed using Theorem~\ref{theorem_generalcase}. The horizontal green line shows the asymptote at $b'_1=\infty$ given by Theorem~\ref{theorem_nbthin}. The vertical red line displays where $b'_1 = b_1$. As expected, both the empirical and theoretical correlations are $0$ at this point. }
\label{fig_corvar}
\end{figure}

Unlike in Theorem~\ref{theorem_nb_binom_thin}, the folds of data that result from applying Algorithm~\ref{alg_gcs} with arbitrary values for $b_j'$ do not necessarily follow negative binomial distributions. The first statement of Theorem~\ref{theorem_generalcase} says that, regardless of the value of $b'_j$ used, the expected value of $\bxm$ and $\bxmm$ are scaled by $\eps_m$ and $(1-\eps_m)$ compared to $\E[\bold{X}]$ in \eqref{eq_meanmodel}. Thus, we can estimate the latent space using the training set $\bxmm$, and validate these estimates using the held out set $\xm$. The second statement of Theorem~\ref{theorem_generalcase} tells us that using the wrong value for $b_j'$ affects the variance of $\bxm$. The third statement says that if $b'_j \neq b_j$, then the training and test sets are correlated, and the magnitude of the correlation grows with the magnitude of the discrepancy between $b_j$ and $b_j'$. The result is displayed and empirically confirmed in Figure~\ref{fig_corvar}. In order to use $\bxm$ to validate a model fit to $\bxmm$ or to do valid inference on latent variables fit to $\bxmm$, we need independence between $\bxm$ and $\bxmm$. Thus, the major takeaway from Theorem~\ref{theorem_generalcase} is that, when the true $b_j$ are unknown, it is important to estimate them well. In Section~\ref{section_simulation_nb}, we use the well-known R package \texttt{sctransform} \citep{hafemeister2019normalization} to estimate each $b_j$.

\subsection{The role of the parameters $\epsilon_1,\ldots,\epsilon_M$}

In this section, we consider the case where we used the ``correct" value of $b_j'$ and thus the results of Theorem~\ref{theorem_nbthin} hold. In this setting, it is simple to show that, for a given fold $m$, the parameter $\epsilon_m$ governs a tradeoff between the amount of information in the training set $\bxmm$ and in the test set $\bxm$. This result is summarized in the following theorem.

\begin{theorem}[Information tradeoff as we vary $\eps$]
\label{theorem_fisher}
If $\bold{X}_{ij} \sim \NB(\mu_{ij}, b_j)$, then the Fisher information contained in a single datapoint $X_{ij}$ for the parameter $\mu_{ij}$ is $I_{\mu_{ij}}(\bx_{ij}) = \frac{b_j}{(b_j + \mu_{ij})\mu_{ij}}$. If we apply Algorithm~\ref{alg_gcs} with $b'_j=b_j$, then for $m=1,\ldots,M$, the Fisher information contained in $\bold{X}^{(m)}_{ij}$ for the parameter $\mu_{ij}$ is $\epsilon_m I_{\mu_{ij}}(\bx_{ij})$, and the corresponding Fisher information contained in $\bold{X}^{(-m)}_{ij}$ is $(1-\epsilon_m) I_{\mu_{ij}}(\bx_{ij})$. 
\end{theorem}

We will see in Section~\ref{section_simulation_nb} that the ideal choice of $\epsilon_1,\ldots,\epsilon_M$ depends on the application.

\section{Simulation Study}
\label{section_simulation_nb}

In this section, we apply negative binomial count splitting to Example 1 and Example 2 from Section~\ref{section_intro_nb}, instantiated to the setting where there are $K^*$ true discrete latent variables which represent cell types that we wish to estimate using $k$-means clustering. More specifically, we consider the following settings:

\begin{list}{}{}
\item{\textbf{Example 1, instantiated to clustering:}} We fit models with $K$ clusters for $K=1,\ldots,10$, evaluate each model, then let the best value of $K$ be our estimate of $K^*$. 
\item{\textbf{Example 2, instantiated to clustering:}} We cluster the data into $K=2$ estimated cell types and test each gene for differential expression across the two estimated clusters.
\end{list}

After introducing our simulation setup in Section~\ref{subsec_datagen}, we show that negative binomial count splitting can be easily applied to Example 1 (Section~\ref{subsec_estK}) and Example 2 (Section~\ref{subsec_difExp}). 

\subsection{Data generating mechanism}
\label{subsec_datagen}

We generate datasets with $n$ cells, $p$ genes, and $K^*$ true cell types. Each gene has a baseline expression level $\exp(\beta_{j0})$ where $\beta_{j0} \overset{\mathrm{i.i.d.}}{\sim} N(0,1)$ for $j=1,\ldots,p$. 

For each dataset, we assign each cell to one of the $K^*$ clusters with equal probability. The first column of the latent variable matrix $L \in \mathbb{R}^{n \times K^*}$ contains ones, and the remaining columns are indicators for clusters $k=2,\ldots,K^*$. The matrix $\beta \in \mathbb{R}^{p \times K^*}$ stores $\beta_{10},\ldots,\beta_{p0}$ in the first column. When $K^* >1$, $\beta_{j2} = \beta^*$ for the first $5\%$ of the genes, and the rest of the entries in the second column of $\beta$ are $0$. If $K^*>2$, then $\beta_{j3} = \beta^*$ for the next $5\%$ of the genes, and all other entries are $0$. We continue filling in the $\beta$ matrix in this manner until all $K^*$ columns have been filled. We consider different values of $\beta^*$ for different datasets, but within a dataset we always use the same value of $\beta^*$ such that all $K^*$ clusters are equally easy to detect. Finally, we let $\log\left(\Lambda\right) = L \beta^\top$. 

We let the overdispersion parameter $b_j$ for the $j$th gene be a function of the average expression $\bar{\Lambda}_j = \frac{1}{n} \sum_{i=1}^n \Lambda_{ij}$ for that gene \citep{choudhary2022comparison, hafemeister2019normalization, love2014moderated}. More specifically, we set $b_j = \frac{\bar{\Lambda}_j}{\tau}$ for either $\tau=1$ (``mild overdispersion") or $\tau=5$ (``severe overdispersion"). We then let $\bx_{ij} \sim \mathrm{NB}(\Lambda_{ij}, b_j)$, such that $\Var(\bold{X}_{ij}) = \Lambda_{ij}\left( 1 + \frac{\Lambda_{ij}}{b_j}\right) \approx \Lambda_{ij}\left( 1 + \tau\right)$. We omit size factors from this simulation study, as they are not the focus of this paper. 

\subsection{Selecting the number of clusters}
\label{subsec_estK}

\subsubsection{Methods}
\label{subsubsec_numclustmethods}

We now introduce the general algorithm used in this section. 

\begin{algorithm}[Estimating the number of clusters] 
Start with datasets $\xtr \in \mathbb{Z}_{\geq 0}^{n \times p}$ and $\xte \in \mathbb{Z}_{\geq 0}^{n \times p}$ and parameter $\eps \in (0,1)$, where $\E\left[ \xte \right] = \frac{1-\eps}{\eps} \E[\xtr]$.  For $K=1,\ldots,10$:
\begin{enumerate}
\item Run k-means clustering to estimate $K$ clusters on $\log(\xtr+1)$. This yields a cluster assignment $\hat{c}_i \in \{1,\ldots,K\}$ for $i = 1,\ldots, n$.
\item For $i=1,\ldots,n$ and $j=1,\ldots,p$, estimate $\E[\bold{X}_{ij}^\mathrm{train}]$ with the sample mean of the training set points assigned to the same cluster:  
$\hat{\mu}^{\mathrm{train}}_{ij} = \frac{1}{\sum_{i'=1}^n \bold{1}\{\hat{c}_{i'} = \hat{c}_i\}} \sum_{i'  = 1}^n X^\mathrm{train}_{i'j} \bold{1}\{\hat{c}_{i'} = \hat{c}_i\}$. 
\item Estimate $\E[\bold{X}_{ij}^\mathrm{test}]$ as follows:
$\hat{\mu}^{\mathrm{test}}_{ij} = \frac{1-\eps}{\eps} \hat{\mu}^{\mathrm{train}}_{ij}$. 
\item Compute the within-cluster mean squared error on the test set as follows: 
\begin{equation}
\label{eq_mse}
MSE(K) = \frac{1}{n \times p} \sum_{i=1}^n \sum_{j=1}^p \left( \log\left(X^\mathrm{test}_{ij}+1\right) - \log\left( \hat{\mu}^{\mathrm{test}}_{ij} +1 \right) \right)^2. 
\end{equation}
\end{enumerate}
\label{alg_estK}
\end{algorithm}

We apply Algorithm~\ref{alg_estK} as follows.  

\begin{list}{}{}
\item{\textbf{Naive method:}} Run Algorithm~\ref{alg_estK} with $\xtr = \xte = X$ and $\eps = 0.5$. 
\item{\textbf{Poisson count splitting (PCS):}} Apply Algorithm~\ref{alg_cs} to $X$ with $M=2$ and some $(\epsilon_1, \epsilon_2)$ to obtain $\xo$ and $\xt$. Run Algorithm~\ref{alg_estK} with $\xtr = \xo, \xte = \xt$, and $\epsilon = \epsilon_1$.  
\item{\textbf{Negative binomial count splitting, known $b$ (NBCS-known):}} Apply Algorithm~\ref{alg_gcs}  to $X$ with $M=2$, some choice of $(\epsilon_1, \epsilon_2)$, and $(b_1', \ldots, b_p')=(b_1,\ldots, b_p)$ to obtain $\xo$ and $\xt$. Run Algorithm~\ref{alg_estK} using $\xtr = \xo, \xte = \xt$, and $\epsilon = \epsilon_1$.

\item{\textbf{Negative binomial count splitting, estimated $b$ (NBCS-estimated):}} First, use the \texttt{R} package \texttt{sctransform} \citep{hafemeister2019normalization} to obtain estimates $\hat{b}_1, \ldots, \hat{b}_p$ of $b_1,\ldots,b_p$. Details are given in Appendix~\ref{appendix_sct}. Then run Algorithm~\ref{alg_gcs} on matrix $X$ with $M=2$, some choice of $(\epsilon_1, \epsilon_2)$, and $(b_1', \ldots, b_p')=(\hat{b}_1,\ldots, \hat{b}_p)$ to obtain $\xo$ and $\xt$. Then run Algorithm~\ref{alg_estK} using $\xtr = \xo, \xte = \xt$, and $\epsilon = \epsilon_1$.
\end{list}{}{}
We now extend both versions of NBCS to perform cross-validation.
\begin{list}{}{}
\item{\textbf{Negative binomial cross-validation, known $b$ (NBCV-known):}} Obtain $(X^{(1)}, \ldots, X^{(M)})$ by running Algorithm~\ref{alg_gcs} on $X$ with $M=10$, $\epsilon_m = \frac{1}{M}$ for $m = 1,\ldots, M$, and $(b_1', \ldots, b_p')=(b_1,\ldots, b_p)$. For $m = 1,\ldots,M$, apply Algorithm~\ref{alg_estK} with 
$\xtr=\xmm$, $\xte=\xm$, and $\eps = \frac{M-1}{M}$. For each value of $K$, record the total MSE summed across the $M$ folds. 
\item{\textbf{Negative binomial cross-validation, estimated $b$ (NBCV-estimated): }} First obtain estimates $\hat{b}_1,\ldots,\hat{b}_p$ of $b_1,\ldots,b_p$ using \texttt{sctransform} on the full data $X$. Then proceed as in NBCV-known, but apply Algorithm~\ref{alg_gcs} with $(b_1', \ldots, b_p') = (\hat{b}_1,\ldots,\hat{b}_p)$.
\end{list}

We already showed in Section~\ref{section_intro_nb} that the naive method fails to provide a viable solution to Examples 1 and 2; nonetheless we include it here for the sake of comparison. We do not consider sample splitting in this section, as we already saw in Section~\ref{section_intro_nb} that it fails to provide a viable solution to our problems of interest. 

\subsubsection{Results}
\label{subsubsec_numclustres}

\begin{figure}
\centering
\includegraphics[width=0.8\textwidth]{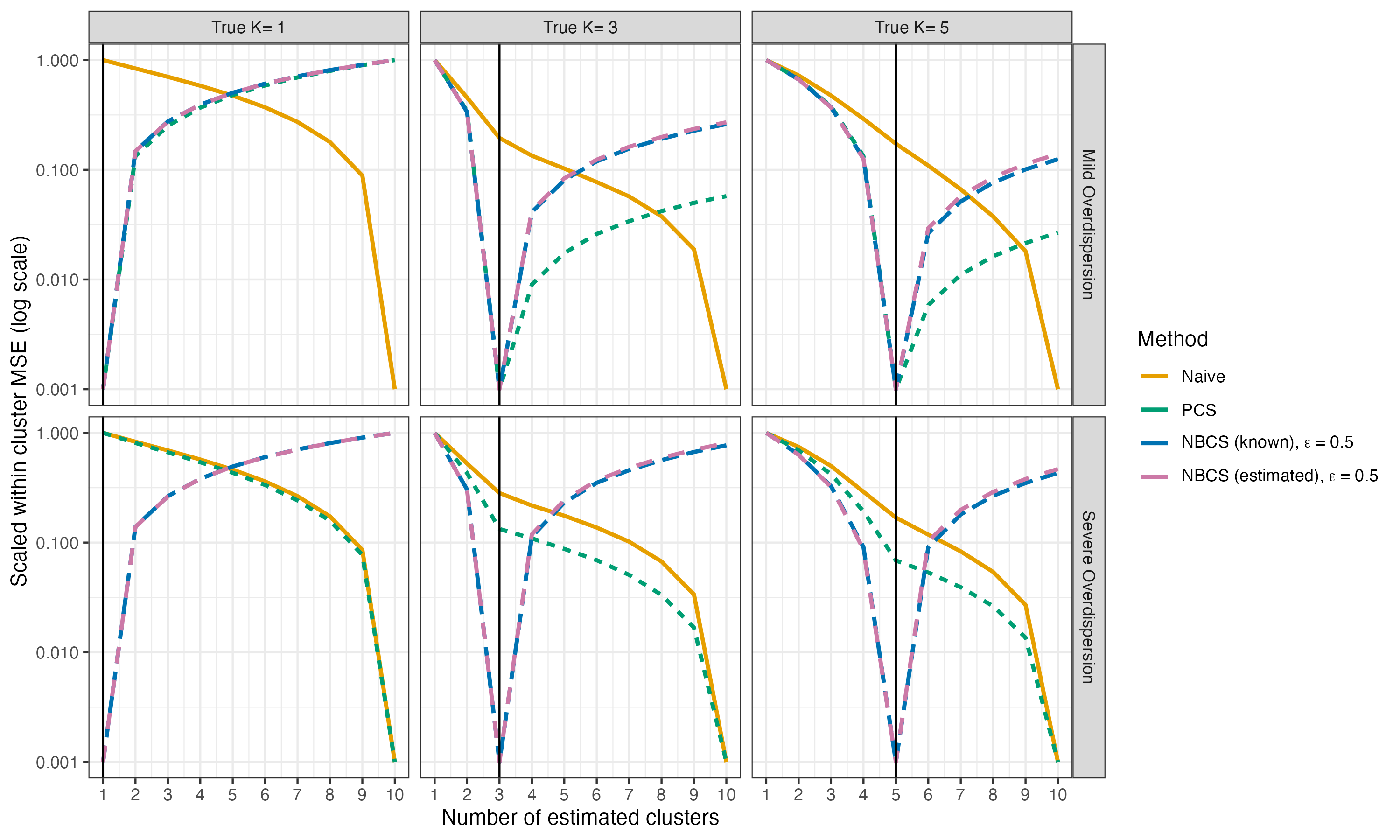}
\caption{The average within-cluster MSE	 (see \eqref{eq_mse} in Algorithm~\ref{alg_estK}) over 1000 datasets for four of the methods described in Section~\ref{subsubsec_numclustmethods}, scaled to have range between 0 and 1.}
\label{fig_mse}
\end{figure}

We first generate 1,000 datasets with $n = 1000$ and $p = 1000$ and  $\beta^* = 1.5$ for $K^* \in \{1,3,5\}$ and the two overdispersion settings described in Section~\ref{subsec_datagen}. For each dataset, we consider the naive method, Poisson count splitting, NBCS-known, and NBCS-estimated (all with $\eps = 0.5$). We plot the average MSE over the 1000 datasets, defined in \eqref{eq_mse}, as a function of $K$. To facilitate comparisons between methods, the y-axis of Figure~\ref{fig_mse} has been scaled such that the MSE for each method ranges from $0$ to $1$.

Figure~\ref{fig_mse} shows that, regardless of the true value of $K^*$ or the amount of overdispersion, the MSE for the naive method decreases monotonically with $K$. Thus, we cannot simply select the value of $K$ that minimizes the loss function. We must instead search for a bend or an ``elbow" in the MSE plot, which fails to provide a clear answer for the number of clusters we should select. We see that PCS performs well (the loss function is minimized when $K^*=K$) under mild overdispersion, but under severe overdispersion its performance approaches that of the naive method. This is as expected from Theorem~\ref{theorem_nb_binom_thin}, since the correlation between $\xtr = X^{(1)}$ and $\xte = X^{(-1)}$ under PCS increases as the amount of overdispersion increases. Both versions of NBCS have loss functions that are minimized when $K=K^*$, regardless of the true value of $K^*$ or the amount of overdispersion in the data. Thus, we can select the number of clusters by selecting the value of $K$ that minimizes the loss function. As the naive method and PCS do not lead to loss functions that are minimized at the true number of clusters, we do not consider them for the remainder of this section. 

\begin{figure}
\centering
\includegraphics[width=\textwidth]{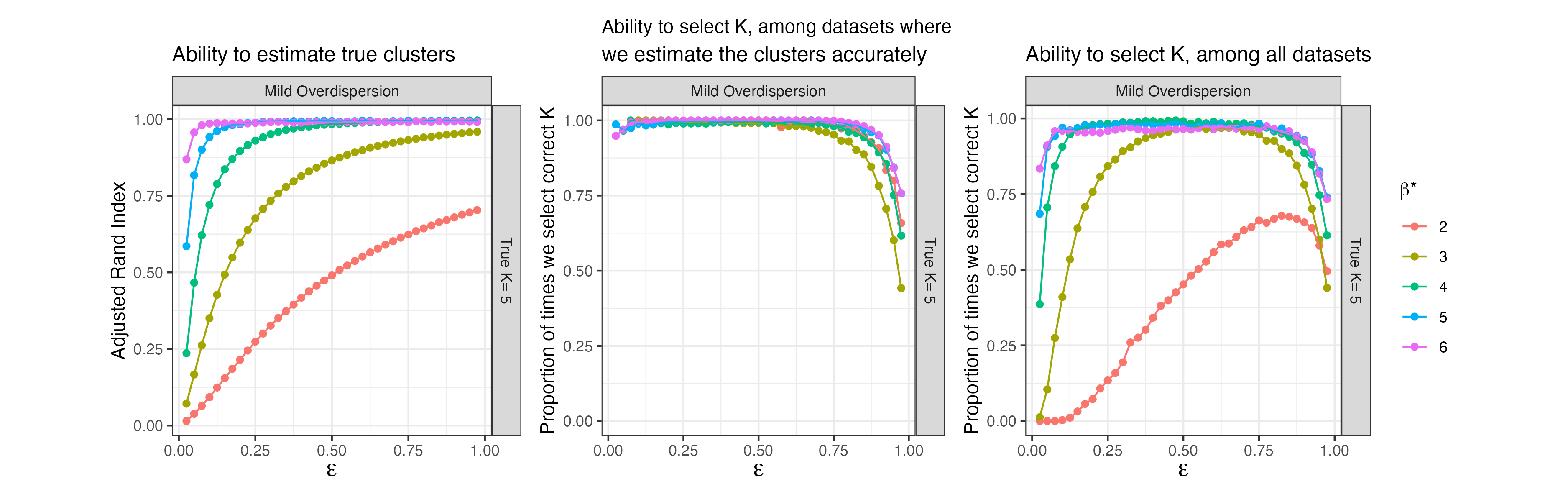}
\caption{ We generate 2,000 datasets with mild overdispersion, $n= 500$, $p=40$, $K^*=5$ for each of five values of $\beta^*$ ranging from $2$ to $6$. For values of $\epsilon$ ranging from $0$ to $1$, we perform NBCS-known. 
\emph{Left: } The average adjusted Rand index between the true clusters and those estimated with on the training set when $K=K^*$, as a function of $\epsilon$. 
\emph{Center: } The proportion of datasets for which the test set MSE is minimized at $K=K^*$, only considering datasets for which the adjusted Rand index between the true clusters and the clusters estimated on the training set when $K=K^*$ exceeds 0.8.
\emph{Right: }  The overall proportion of datasets for which the loss function is minimized at $K=K^*$.}
\label{fig_roleEps}
\end{figure}

We next explore the role of the parameter $\epsilon$ in NBCS. We generate 2,000 datasets with mild overdispersion with $n= 500$ and $p=40$ for $K=5$ and for $\beta^*$ values ranging from $2$ to $6$. For values of $\epsilon$ ranging from $0$ to $1$, we perform NBCS-known. For each dataset and each value of $\epsilon$, we consider three metrics. The left panel of Figure~\ref{fig_roleEps} displays the adjusted Rand index (ARI) between the true clusters and those estimated using $\xtr$ when $K=K^*$, as a function of $\epsilon$. For a given signal strength, the average ARI increases with $\epsilon$ because a large value of $\eps$ means that we use more of the information in our data in the cluster estimation phase of Algorithm~\ref{alg_estK} (see Theorem~\ref{theorem_fisher}). The center panel of Figure~\ref{fig_roleEps} shows the proportion of times that the MSE is minimized at $K=K^*$ (i.e. that we select the correct value of $K$), given that the ARI between the true clusters and the estimated clusters when $K=K^*$ exceeds $0.8$. This metric decreases with $\epsilon$, as large values of $\eps$ leave less information in the test set for us to validate the estimated clusters. The right panel of Figure~\ref{fig_roleEps} shows the overall proportion of datasets for which the loss function is minimized at $K=K^*$. The optimal value of $\eps$ depends on the true signal strength, but it always involves a tradeoff between choosing $\eps$ large enough to estimate good clusters on $\xtr$, but not so large that we cannot accurately validate the clusters. 

\begin{figure}
\includegraphics[width=\textwidth]{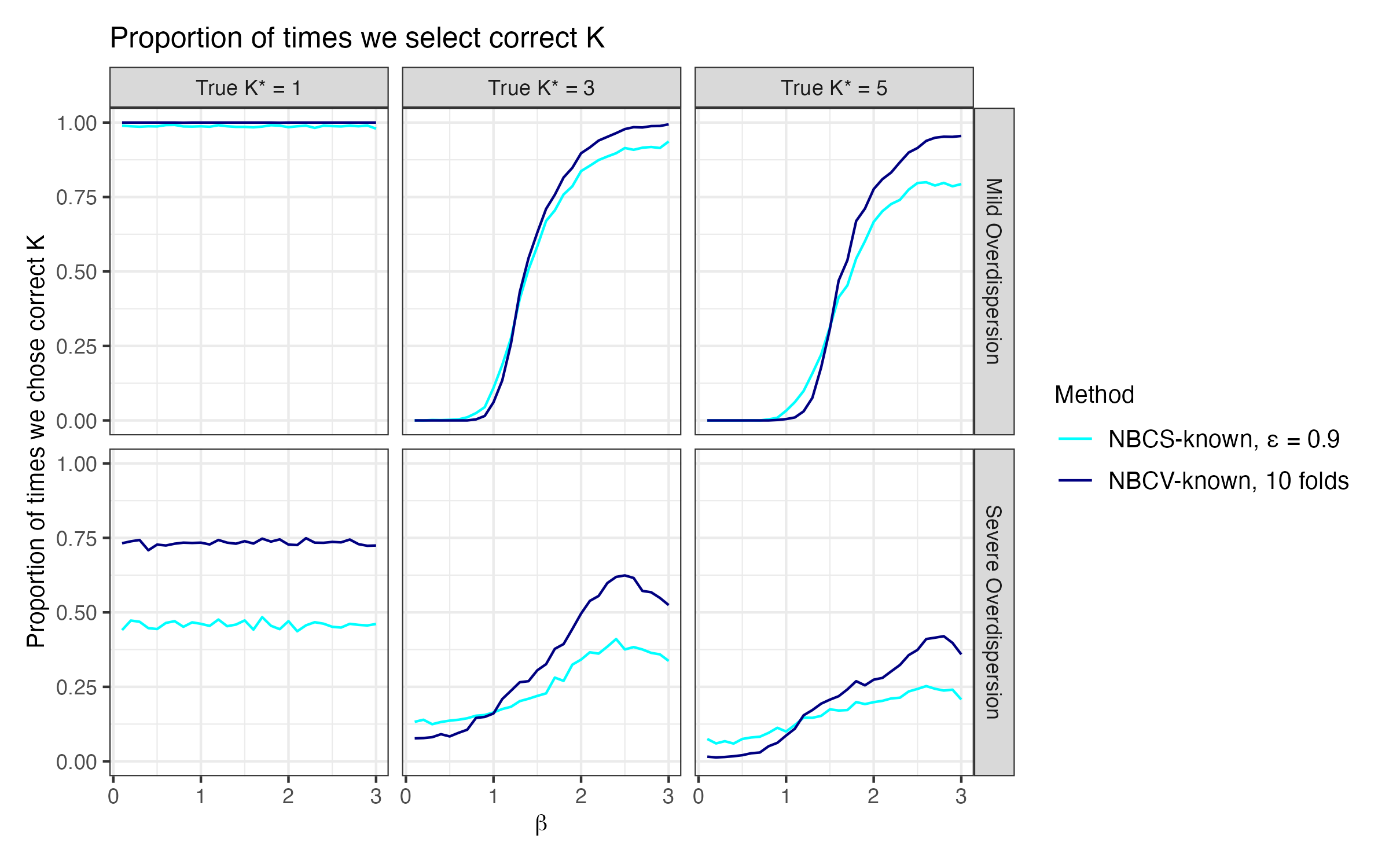}
\caption{We fix $n=500$ and $p=40$ and we generate 200 datasets for each combination of $K \in \{1,3,5\}$, both overdispersion settings, and $\beta^*$ values ranging from $0.1$ to $3$. For each value of $\beta^*$, we show the proportion of datasets for which the loss function was minimized at $K=K^*$. }
\label{fig_propCor}
\end{figure}

Finally, we compare NBCS-known with $\eps = 0.9$ to NBCV-known with 10 folds. We generate 2,000 datasets where $n=500$ and $p=40$ for values of $\beta^*$ ranging from $1$ to $6$ and $K^* = 1,3,5$. As both methods use 90\% of the information in the data for training and 10\% for testing, these methods have the same average MSE curves over many datasets. However, for a given dataset, NBCV-known selects the correct value of $K$ more often than NBCS-known, because averaging the MSE over 10 folds reduces the variance in our validation step (Figure~\ref{fig_propCor}). 

\subsection{Testing for differential expression}
\label{subsec_difExp}

\subsubsection{Methods}

In this section we let $K^*=2$ and we always estimate two clusters on the data. Our focus is no longer on estimating the number of clusters, but rather on studying differential expression across a given set of estimated clusters. We use the following algorithm. 

\begin{algorithm}[Testing for differential expression.] Start with datasets $\xtr \in \mathbb{Z}_{\geq 0}^{n \times p}$ and $\xte \in \mathbb{Z}_{\geq 0}^{n \times p}$. 
\begin{enumerate}	
\item Apply $k$-means clustering with $K=2$ to estimate clusters on $\log(\xtr+1)$. This yields a cluster assignment $\hat{c}_i \in \{0,1\}$ for $i=1,\ldots,n$. 
\item For $j=1,\ldots,p$, fit a negative binomial GLM of $X^\mathrm{test}_j$ on $\hat{c}$. Report the Wald p-value for the slope coefficient. 
\end{enumerate}
\label{alg_diffExp}
\end{algorithm}

We consider the following ways to obtain $\xtr$ and $\xte$ in Algorithm~\ref{alg_diffExp}. 

\begin{list}{}{}
\item{\textbf{Naive method:}} Let $\xtr = \xte = X$.
\item{\textbf{Poisson count splitting (PCS):}} Obtain $\xtr = X^{(1)}$ and $\xte = X^{(2)}$ by running Algorithm~\ref{alg_cs} on the data $X$ with $M=2$ and $(\epsilon_1, \epsilon_2) = (\epsilon, 1-\epsilon)$. 
\item{\textbf{Negative binomial count splitting, known $b$ (NBCS-known):}} Obtain $\xtr = \xo$ and $\xte= X^{(2)}$ by running Algorithm~\ref{alg_gcs} on $X$ with $M=2$, $(\epsilon_1, \epsilon_2) = (\epsilon, 1-\epsilon)$, and $(b_1', \ldots, b_p')=(b_1,\ldots, b_p)$. 
\item{\textbf{Negative binomial count splitting, estimated $b$ (NBCS-estimated:)}} Use the \texttt{R} package \texttt{sctransform} \citep{hafemeister2019normalization} to obtain estimates $\hat{b}_1, \ldots, \hat{b}_p$ of $b_1,\ldots,b_p$ using the full dataset $X$. Then apply Algorithm~\ref{alg_gcs} on matrix $X$ with $M=2$, $(\epsilon_1, \epsilon_2) = (\epsilon, 1-\epsilon)$, and $(b_1', \ldots, b_p')=(\hat{b}_1,\ldots, \hat{b}_p)$. 
\end{list}

Unlike in Section~\ref{subsec_estK}, we do not consider splitting with $M>2$ folds and we do not aggregate results across folds by interchanging the roles of the train and test sets. We leave the possibility of aggregating differential expression test statistics across multiple folds to future work. Once again, we do not consider sample splitting because we already saw in Figure~\ref{fig_intro} that it fails to provide a viable solution to in Example~\ref{ex:diffExp}. 

\subsubsection{Results}

\begin{figure}
\centering
\includegraphics[width=0.8\textwidth]{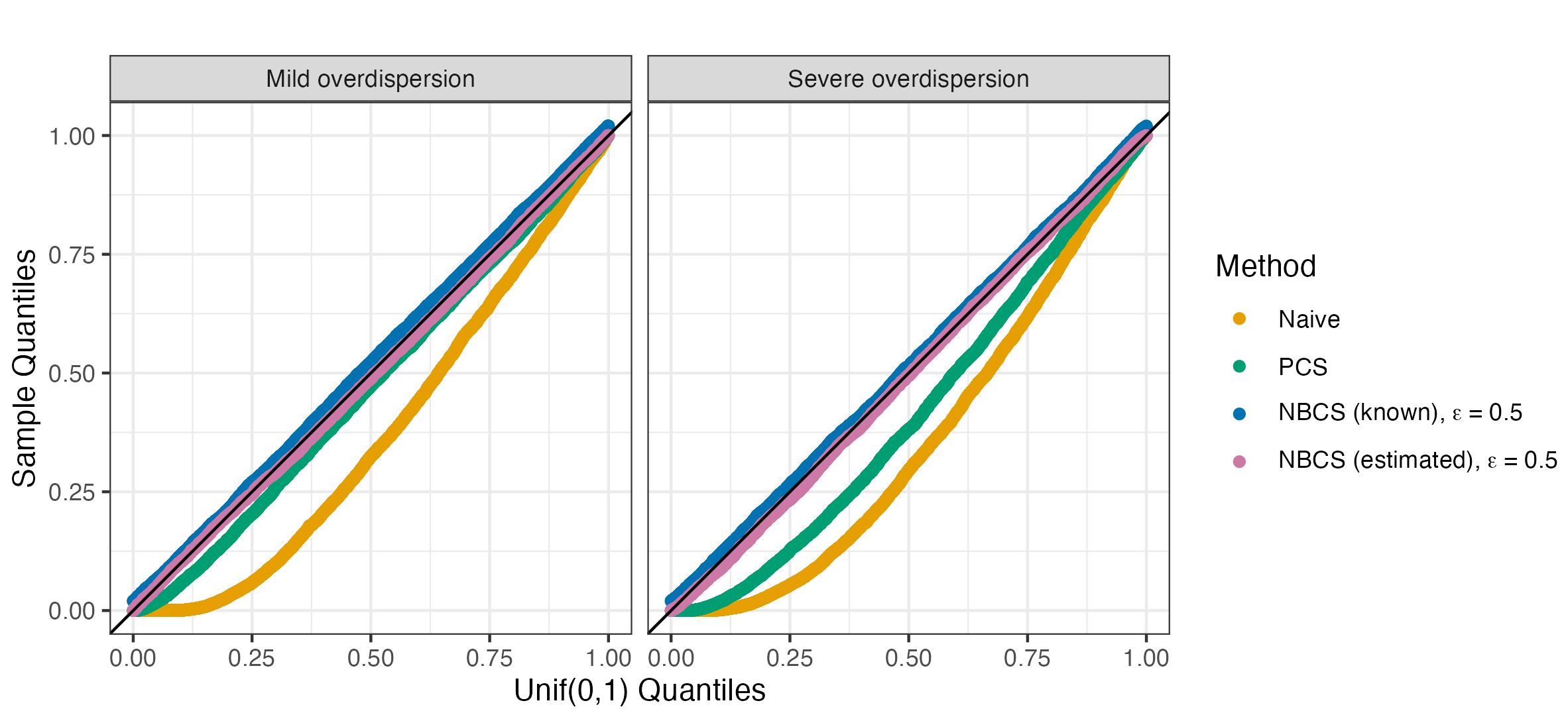}
\caption{We generate $1,000$ datasets with $K^*=2, n=500,$ and $p=40$ for each overdispersion setting and for each of 16 values of $\beta^*$ ranging from $0$ to $3$. For each dataset, we carry out the four variations of Algorithm~\ref{alg_diffExp} given in Section~\ref{subsec_difExp}. The QQ-plots indicate that only NBCS controls the Type 1 error rate across both overdispersion settings. }
\label{fig_QQ}
\end{figure}

We generate datasets using the mechanism described in Section~\ref{subsec_datagen} with $K^*=2$, $n=500$, and $p=40$. Under this mechanism, the first two genes are differentially expressed across the two true clusters. We refer to the remaining $38$ genes as null genes because they have the same expected expression across all cells. Figure~\ref{fig_QQ} shows uniform QQ plots of the p-values obtained from the four variations of Algorithm~\ref{alg_diffExp} for the null genes, aggregated across 1000 datasets for each of 16 $\beta^*$ values. We see from Figure~\ref{fig_QQ} that both the naive method and Poisson count splitting fail to control the Type 1 error rate. Poisson count splitting performs worse when overdispersion is severe. On the other hand, both versions of NBCS control the Type 1 error rate. 

\begin{figure}
\centering
\includegraphics[width=\textwidth]{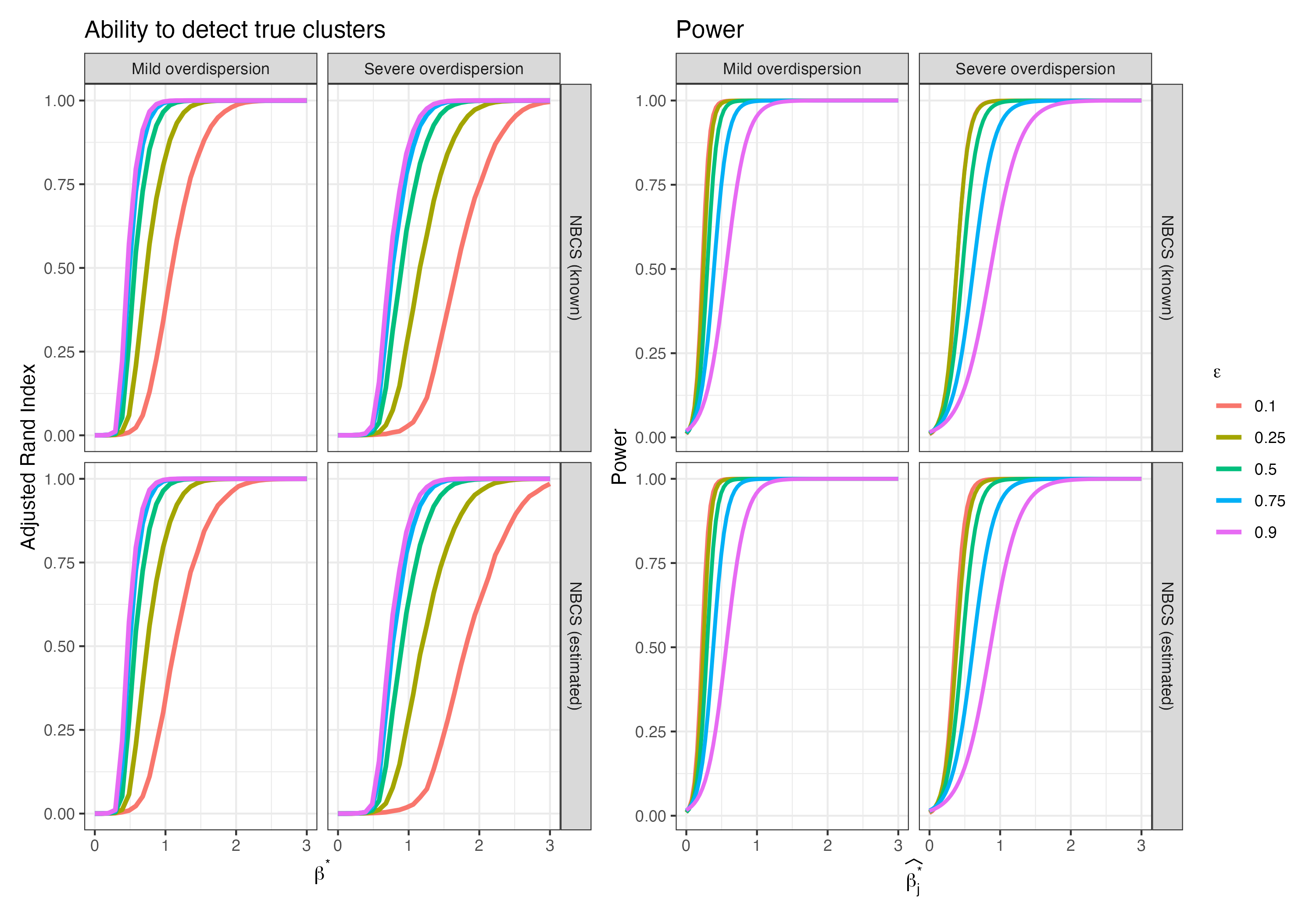}
\caption{\emph{Left:} The average adjusted Rand index between the true clusters and those estimated on the training set, plotted as a function of $\beta^*$ for both overdispersion settings and for each value of $\epsilon$. \emph{Right:} The proportion of times that the differential expression p-value for a non-null gene was less than $0.05$, as a function of the association between the gene's expected expression and the estimated clusters.}
\label{fig_detectPower}
\end{figure}

Finally, we explore the role of $\eps$ in this setting. We generate $1000$ datasets where $n=1000$ and $p=1000$ for each of 16 values of $\beta^*$ ranging from $0$ to $3$. For each dataset, we consider the adjusted Rand index between the true clusters and those estimated on the training set. As in Section~\ref{subsec_estK}, we expect that larger values of $\epsilon$ will lead to higher adjusted Rand indices, on average, for a given signal strength $\beta^*$. This is confirmed in the left panel of Figure~\ref{fig_detectPower}. On the other hand, given the clusters that we estimated on the training set, smaller values of $\epsilon$ leave us with more power to detect differential expression on the test set. We define $\hat{\beta}_j^*$ to be the estimated GLM coefficient for a gene $X_j$ if we regress its mean vector $\Lambda_j$ onto the clusters estimated on the training set; note that $\hat{\beta}_j^* = \beta^*$ only if the estimated clusters are exactly equal to the true clusters. The right panel of Figure~\ref{fig_detectPower} plots the proportion of times that the differential expression p-value for a non-null gene was less than $0.05$, as a function of $\hat{\beta}_j^*$. We see that, for a given value of $\hat{\beta}_j^*$, the proportion of null hypotheses rejected is highest when $\epsilon$ is small, because smaller values of $\epsilon$ leave more information in the test set. 

\section{Application to fetal cell atlas data}
\label{section_realData_nb}

In this section, we apply negative binomial count splitting to solve the problems associated with Example~\ref{ex:stable} from Section~\ref{section_intro_nb}. \cite{cao2020human} sequenced more than 4 million cells from 121 human fetal samples to create a fetal cell atlas, with a goal of organizing the cells from each of 15 different organs into cell types and cell subtypes. In one analysis, after preprocessing, they clustered the expression data from 
the kidney cells into nine main cell types. They then subsetted the data to include only the cluster thought to correspond to metanephric kidney cells, and clustered that subset into 10 cell subtypes. As there is no ground truth available for the main cell types or the cell subtypes, it is difficult to validate their results. \cite{cao2020human} use a procedure that they call \emph{intradataset cross-validation}; see their Figure 2. We describe a simplified version of this procedure in Algorithm~\ref{alg_cao_intra}. 

\begin{algorithm}[Intradataset cross-validation, \cite{cao2020human}]
\label{alg_cao_intra} 
Input: a cell-by-gene expression matrix $X \in \mathbb{Z}_{\geq 0}^{n \times p}$. 
\begin{enumerate}
\item After preprocessing, cluster the dataset $X$ to obtain estimated cell types $\hat{L}(X)^\mathrm{cluster}$. 
\item Divide the $n$ cells into $5$ folds and, for $m = 1,\ldots, 5$: 
\begin{enumerate}
\item Let all cells in the $m$th fold be the test set; the remaining cells are the training set. 
\item Train a classifier to predict $\hat{L}(X)^\mathrm{cluster}_i$ using $X_i$, where $i$ indexes the cells in the training set. 
\item Use this trained classifier
to predict $\hat{L}^{\mathrm{cluster}}_{i'}$ using $X_{i'}$, where $i'$ indexes the cells in the test set. Let $\hat{L}_{i'}^\mathrm{classifier}$ denote the prediction for the $i'$th cell. 
\end{enumerate} 
\item For $i=1,\ldots,n$, the $i$th cell now has a value $\hat{L}(X)^\mathrm{cluster}_i$, obtained from clustering, and $\hat{L}^{\mathrm{classifier}}_i$, obtained via prediction when it belonged to the test set. Make a confusion matrix comparing  $\hat{L}(X)^\mathrm{cluster}$ and $\hat{L}^{\mathrm{classifier}}$. A diagonal confusion matrix is treated as evidence that the clusters are reproducible, since the cluster assignment for a given cell can be recovered by a classifier that was trained without knowledge of that cell's cluster assignment. The adjusted Rand index (ARI) numerically summarizes the degree of agreement between $\hat{L}(X)^\mathrm{cluster}$ and $\hat{L}^{\mathrm{classifier}}$.
\end{enumerate}	
\end{algorithm}

In Section~\ref{section_intro_nb} (Figure~\ref{fig_intro}(e)), we showed using a toy dataset that this procedure is problematic. Because the entire dataset $X$ is used in Step 1 to estimate the clusters $\hat{L}(X)^\mathrm{cluster}$, the test set in Step 2(c) has not truly been held out of the training process.  Thus, despite the fact that the clusters estimated on the toy dataset are driven by random noise, the confusion matrix output by Algorithm~\ref{alg_cao_intra} is close to diagonal and the ARI is close to $1$ (Figure~\ref{fig_intro}(e)). Negative binomial count splitting provides a simple alternative (Figure~\ref{fig_intro}(f)), which we outline below. 

\begin{algorithm}[Intradataset cross-validation via count splitting]
\label{alg_cs_intra} \hspace{1mm} \\
Input: a cell-by-gene expression matrix $X \in \mathbb{Z}_{\geq 0}^{n \times p}$. 
\begin{enumerate}
\item Obtain an estimate $\hat{b}_j$ of the gene-specific overdispersion parameter $b_j$ for $j=1,\ldots,p$. 
\item Apply Algorithm~\ref{alg_gcs} (negative binomial count splitting) with $M=2$, $\epsilon_1 = \epsilon_2 = 0.5$, and $(b_1',\ldots, b_p') = (\hat{b}_1,\ldots,\hat{b}_p)$ to create two folds of data, $\xo$ and $\xt$.
\item After preprocessing, apply a clustering algorithm to $\xo$ to obtain an estimated cluster $\hat{L}(X^{(1)})_i$ for the $i$th cell for $i=1,\ldots,n$. 
\item After preprocessing, apply the same clustering algorithm to $\xt$ to obtain an estimated cluster $\hat{L}(X^{(2)})_i$ for the $i$th cell for $i=1,\ldots,n$. 
\item Make a confusion matrix comparing $\hat{L}(X^{(1)})$ and $\hat{L}(X^{(2)})$. A diagonal confusion matrix (up to a permutation of the columns) is treated as evidence of cluster reproducibility, since the cells are reliably assigned to the same cluster on independent realizations of the data. The ARI numerically summarizes the degree of agreement between $\hat{L}(X^{(1)})$ and $\hat{L}(X^{(2)})$. 
\end{enumerate}	
\end{algorithm}

We now compare Algorithm~\ref{alg_cao_intra} and Algorithm~\ref{alg_cs_intra} on data from the human fetal cell atlas. We consider two versions of Algorithm~\ref{alg_cs_intra}. We apply a version where, in Step 1, we let each $\hat{b}_j = \infty$, which corresponds to assuming that the data are Poisson. We also apply a version where, in Step 1, we estimate each $\hat{b}_j$ using the full dataset using \texttt{sctransform}; details are given in Appendix~\ref{appendix_sct}. In both cases, we use $\epsilon_1=\epsilon_2=0.5$ such that $\xo$ and $\xt$ are identically distributed. Our goal is not to discover the optimal clusters in this dataset, but rather to compare Algorithm~\ref{alg_cao_intra} and Algorithm~\ref{alg_cs_intra} as strategies for validating clusters. As such, we do not attempt to reproduce the exact analysis from \cite{cao2020human}; we use a simplified implementation, which is described in Appendix~\ref{appendix_cao}.

\begin{figure}[h]
\includegraphics[width=\textwidth]{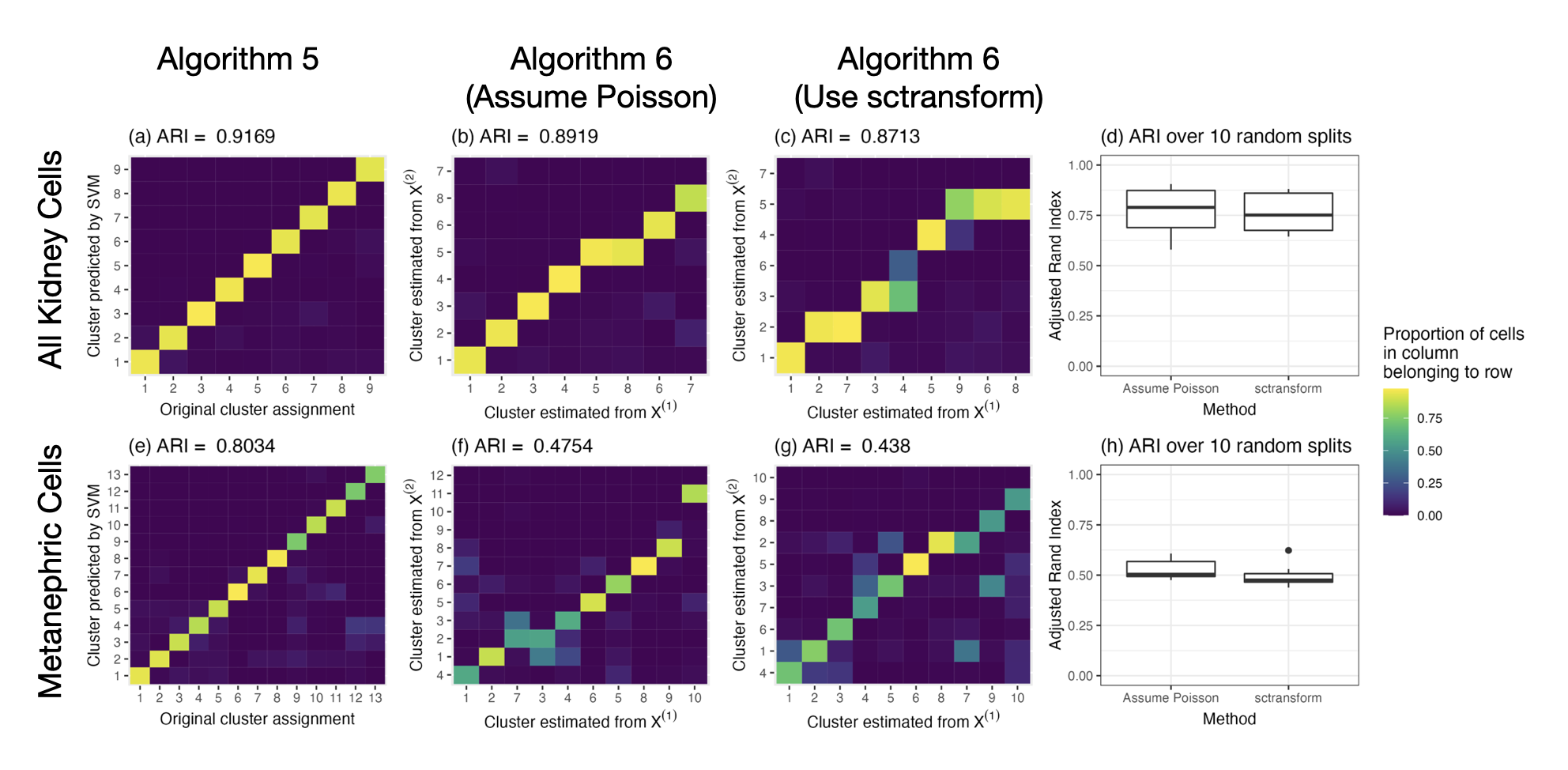}
\caption{The results of applying (a) Algorithm~\ref{alg_cao_intra}, (b) Algorithm~\ref{alg_cs_intra} assuming the data is Poisson, and (c) Algorithm~\ref{alg_cs_intra} with overdispersion parameters estimated via \texttt{sctransform}, to the full kidney dataset. Panels (d), (e), and (f) show the results of applying those algorithms to only the metanephric cells. Panels (d) and (h) show the adjusted Rand indices (ARIs) that result from 10 replications of Algorithm~\ref{alg_cs_intra}.}
\label{fig_realData_confusion}	
\end{figure}

To start, we apply Algorithm~\ref{alg_cao_intra} and both versions of Algorithm~\ref{alg_cs_intra} to all $178,603$ kidney cells from the human fetal cell atlas. The results are shown in panels (a), (b), and (c) of Figure~\ref{fig_realData_confusion}. Regardless of the algorithm used, we see diagonal confusion matrices and high ARIs, suggesting reproducibility of the clusters. For panels (b) and (c), we permute the columns of the matrices to make them appear as diagonal as possible, but we note that the ARI is invariant to these permutations. 

We next apply Algorithm~\ref{alg_cao_intra} and both versions of Algorithm~\ref{alg_cs_intra} to the $90,876$ kidney cells that \cite{cao2020human} annotated as metanephric cells. The results are shown in  panels (d), (e), and (f) of Figure~\ref{fig_realData_confusion}. Overall, panels (e) and (f) show lower ARIs than panels (b) and (c), suggesting that these supposed cell subtypes are somewhat less reproducible than the main cell types. We note that the difference between the main cell type analysis and the cell subtype analysis is least stark for Algorithm~\ref{alg_cao_intra}, where the double use of data causes the cell subtypes to appear more reproducible than they are in both analyses. Similarly, the difference 
between the main cell type analysis and the cell subtype analysis is less stark for the Poisson version of Algorithm~\ref{alg_cs_intra} than the \texttt{sctransform} version, suggesting that the dependence between the training set and the test set induced by setting $\hat{b}_1 = \ldots = \hat{b}_p = \infty$ is also enough to make the cell subtypes appear slightly more reproducible than they are.

We repeat each version of Algorithm~\ref{alg_cs_intra} ten times, using ten different random splits of the data. The resulting ARIs are shown in Figures~\ref{fig_realData_confusion}(d) and \ref{fig_realData_confusion}(h). The ARIs from the cell subtype analysis are consistently lower than those from the main cell type analysis.

In summary, due to the double use of data, Algorithm~\ref{alg_cao_intra} overestimates the reproducibility of the metanephric cell subtypes. On the other hand, Algorithm~\ref{alg_cs_intra} allows us to see that the metanephric cell subtypes are less reproducible than the main kidney cell types. 

\section{Discussion}
\label{sec_disc}

In this paper, we introduced an algorithm to split a negative binomial random variable into two or more independent negative binomial random variables, and we applied it to the analysis of scRNA-seq data. This algorithm is also applicable in other settings; data are modeled as negative binomial in a wide variety of fields. Furthermore, while we were particularly motivated by unsupervised settings in which sample splitting is not an option, negative binomial count splitting can also be used in supervised settings such as inference after variable selection in negative binomial regression. For further discussion, see \cite{neufeld2023data} and \cite{dharamshi2023generalized}. An implementation of the techniques in this paper is available in the \texttt{R} package \texttt{countsplit}, and tutorials showing how this can be integrated with existing scRNA-seq software packages, are available at \texttt{anna-neufeld.github.io/countsplit.tutorials}.

\section*{Acknowledgements}

Anna Neufeld and Daniela Witten were supported by the Simons Foundation, the National Institutes of Health (NIH), and the Keck Foundation. Lucy Gao was supported by the National Sciences and Engineering Council of Canada. Joshua Popp was supported by the NIH. Alexis Battle was supported by the NIH and the Chan Zuckerberg Initiative.

\bibliographystyle{plain}
\bibliography{nbcs.bib}

\begin{thebibliography}{32}
\providecommand{\natexlab}[1]{#1}
\providecommand{\url}[1]{\texttt{#1}}
\expandafter\ifx\csname urlstyle\endcsname\relax
  \providecommand{\doi}[1]{doi: #1}\else
  \providecommand{\doi}{doi: \begingroup \urlstyle{rm}\Url}\fi

\bibitem[Abdelaal et~al.(2019)Abdelaal, Michielsen, Cats, Hoogduin, Mei,
  Reinders, and Mahfouz]{abdelaal2019comparison}
Tamim Abdelaal, Lieke Michielsen, Davy Cats, Dylan Hoogduin, Hailiang Mei,
  Marcel~JT Reinders, and Ahmed Mahfouz.
\newblock {A comparison of automatic cell identification methods for
  single-cell RNA sequencing data}.
\newblock \emph{Genome Biology}, 20:\penalty0 1--19, 2019.

\bibitem[Aizarani et~al.(2019)Aizarani, Saviano, Mailly, Durand, Herman,
  Pessaux, et~al.]{aizarani2019human}
Nadim Aizarani, Antonio Saviano, Laurent Mailly, Sarah Durand, Josip~S Herman,
  Patrick Pessaux, et~al.
\newblock A human liver cell atlas reveals heterogeneity and epithelial
  progenitors.
\newblock \emph{Nature}, 572\penalty0 (7768):\penalty0 199--204, 2019.

\bibitem[Batson et~al.(2019)Batson, Royer, and Webber]{batson2019molecular}
Joshua Batson, Lo{\"\i}c Royer, and James Webber.
\newblock {Molecular cross-validation for single-cell RNA-seq}.
\newblock \emph{BioRxiv}, page 786269, 2019.

\bibitem[Cao et~al.(2020)Cao, O’Day, Pliner, Kingsley, Deng, Daza,
  et~al.]{cao2020human}
Junyue Cao, Diana~R O’Day, Hannah~A Pliner, Paul~D Kingsley, Mei Deng, Riza~M
  Daza, et~al.
\newblock A human cell atlas of fetal gene expression.
\newblock \emph{Science}, 370\penalty0 (6518):\penalty0 eaba7721, 2020.

\bibitem[Chen and Witten(2023)]{chen2022selective}
Yiqun~T Chen and Daniela~M Witten.
\newblock Selective inference for k-means clustering.
\newblock \emph{Journal of Machine Learning Research}, 2023.

\bibitem[Choudhary and Satija(2022)]{choudhary2022comparison}
Saket Choudhary and Rahul Satija.
\newblock {Comparison and evaluation of statistical error models for
  scRNA-seq}.
\newblock \emph{Genome Biology}, 23\penalty0 (1):\penalty0 1--20, 2022.

\bibitem[Chung and Storey(2015)]{chung2015statistical}
Neo~Christopher Chung and John~D Storey.
\newblock Statistical significance of variables driving systematic variation in
  high-dimensional data.
\newblock \emph{Bioinformatics}, 31\penalty0 (4):\penalty0 545--554, 2015.

\bibitem[Dharamshi et~al.(2023)Dharamshi, Neufeld, Motwani, Gao, Witten, and
  Bien]{dharamshi2023generalized}
Ameer Dharamshi, Anna Neufeld, Keshav Motwani, Lucy~L Gao, Daniela Witten, and
  Jacob Bien.
\newblock Generalized data thinning using sufficient statistics.
\newblock \emph{arXiv preprint arXiv:2303.12931}, 2023.

\bibitem[Durrett(2019)]{durrett2019probability}
Rick Durrett.
\newblock \emph{Probability: Theory and Examples}, volume~49.
\newblock Cambridge University Press, 2019.

\bibitem[Eraslan et~al.(2019)Eraslan, Simon, Mircea, Mueller, and
  Theis]{eraslan2019single}
G{\"o}kcen Eraslan, Lukas~M Simon, Maria Mircea, Nikola~S Mueller, and Fabian~J
  Theis.
\newblock {Single-cell RNA-seq denoising using a deep count autoencoder}.
\newblock \emph{Nature Communications}, 10\penalty0 (1):\penalty0 1--14, 2019.

\bibitem[Fu and Perry(2020)]{fu2020estimating}
Wei Fu and Patrick~O Perry.
\newblock Estimating the number of clusters using cross-validation.
\newblock \emph{Journal of Computational and Graphical Statistics}, 29\penalty0
  (1):\penalty0 162--173, 2020.

\bibitem[Gao et~al.(2022)Gao, Bien, and Witten]{gao2022selective}
Lucy~L Gao, Jacob Bien, and Daniela Witten.
\newblock Selective inference for hierarchical clustering.
\newblock \emph{Journal of the American Statistical Association}, pages 1--11,
  2022.

\bibitem[Grabski et~al.(2023)Grabski, Street, and
  Irizarry]{grabski2022significance}
Isabella~N Grabski, Kelly Street, and Rafael~A Irizarry.
\newblock {Significance analysis for clustering with single-Cell RNA-sequencing
  data}.
\newblock \emph{Nature Methods}, pages 1--7, 2023.

\bibitem[Hafemeister and Satija(2019)]{hafemeister2019normalization}
Christoph Hafemeister and Rahul Satija.
\newblock {Normalization and variance stabilization of single-cell RNA-seq data
  using regularized negative binomial regression}.
\newblock \emph{{Genome Biology}}, 20\penalty0 (1):\penalty0 1--15, 2019.

\bibitem[Harremo{\"e}s et~al.(2010)Harremo{\"e}s, Johnson, and
  Kontoyiannis]{harremoes2010thinning}
Peter Harremo{\"e}s, Oliver Johnson, and Ioannis Kontoyiannis.
\newblock Thinning, entropy, and the law of thin numbers.
\newblock \emph{IEEE Transactions on Information Theory}, 56\penalty0
  (9):\penalty0 4228--4244, 2010.

\bibitem[Hastie et~al.(2009)Hastie, Tibshirani, Friedman, and
  Friedman]{hastie2009elements}
Trevor Hastie, Robert Tibshirani, Jerome~H Friedman, and Jerome~H Friedman.
\newblock \emph{The Elements of Statistical Learning: Data Mining, Inference,
  and Prediction}.
\newblock Springer, 2009.

\bibitem[Hubert and Arabie(1985)]{hubert1985comparing}
Lawrence Hubert and Phipps Arabie.
\newblock Comparing partitions.
\newblock \emph{Journal of Classification}, 2\penalty0 (1):\penalty0 193--218,
  1985.

\bibitem[Joe(1996)]{joe1996time}
Harry Joe.
\newblock Time series models with univariate margins in the convolution-closed
  infinitely divisible class.
\newblock \emph{Journal of Applied Probability}, 33\penalty0 (3):\penalty0
  664--677, 1996.

\bibitem[Lange et~al.(2004)Lange, Roth, Braun, and Buhmann]{lange2004stability}
Tilman Lange, Volker Roth, Mikio~L Braun, and Joachim~M Buhmann.
\newblock Stability-based validation of clustering solutions.
\newblock \emph{Neural Computation}, 16\penalty0 (6):\penalty0 1299--1323,
  2004.

\bibitem[Leiner et~al.(2022)Leiner, Duan, Wasserman, and
  Ramdas]{leiner2021data}
James Leiner, Boyan Duan, Larry Wasserman, and Aaditya Ramdas.
\newblock Data fission: splitting a single data point.
\newblock \emph{arXiv preprint arXiv:2112.11079}, 2022.

\bibitem[Lopez et~al.(2018)Lopez, Regier, Cole, Jordan, and
  Yosef]{lopez2018deep}
Romain Lopez, Jeffrey Regier, Michael~B Cole, Michael~I Jordan, and Nir Yosef.
\newblock Deep generative modeling for single-cell transcriptomics.
\newblock \emph{Nature Methods}, 15\penalty0 (12):\penalty0 1053--1058, 2018.

\bibitem[Love et~al.(2014)Love, Huber, and Anders]{love2014moderated}
Michael~I Love, Wolfgang Huber, and Simon Anders.
\newblock {Moderated estimation of fold change and dispersion for RNA-seq data
  with DESeq2}.
\newblock \emph{{Genome Biology}}, 15\penalty0 (12):\penalty0 1--21, 2014.

\bibitem[McKenzie(1986)]{mckenzie1986autoregressive}
Ed~McKenzie.
\newblock Autoregressive moving-average processes with negative-binomial and
  geometric marginal distributions.
\newblock \emph{Advances in Applied Probability}, 18\penalty0 (3):\penalty0
  679--705, 1986.

\bibitem[Neufeld et~al.(2022)Neufeld, Gao, Popp, Battle, and
  Witten]{neufeld2022inference}
Anna Neufeld, Lucy~L Gao, Joshua Popp, Alexis Battle, and Daniela Witten.
\newblock {Inference after latent variable estimation for single-cell RNA
  sequencing data}.
\newblock \emph{Biostatistics}, 2022.

\bibitem[Neufeld et~al.(2023)Neufeld, Dharamshi, Gao, and
  Witten]{neufeld2023data}
Anna Neufeld, Ameer Dharamshi, Lucy~L Gao, and Daniela Witten.
\newblock Data thinning for convolution-closed distributions.
\newblock \emph{arXiv preprint arXiv:2301.07276}, 2023.

\bibitem[Owen and Perry(2009)]{owen2009bi}
Art~B Owen and Patrick~O Perry.
\newblock Bi-cross-validation of the svd and the nonnegative matrix
  factorization.
\newblock \emph{The Annals of Applied Statistics}, 3\penalty0 (2):\penalty0
  564--594, 2009.

\bibitem[Sarkar and Stephens(2021)]{sarkar2021separating}
Abhishek Sarkar and Matthew Stephens.
\newblock {Separating measurement and expression models clarifies confusion in
  single-cell RNA sequencing analysis}.
\newblock \emph{Nature Genetics}, 53\penalty0 (6):\penalty0 770--777, 2021.

\bibitem[Tibshirani and Walther(2005)]{tibshirani2005cluster}
Robert Tibshirani and Guenther Walther.
\newblock Cluster validation by prediction strength.
\newblock \emph{Journal of Computational and Graphical Statistics}, 14\penalty0
  (3):\penalty0 511--528, 2005.

\bibitem[Townes et~al.(2019)Townes, Hicks, Aryee, and
  Irizarry]{townes2019feature}
F.~W. Townes, S.~C. Hicks, M.~J. Aryee, and R.~A. Irizarry.
\newblock {Feature selection and dimension reduction for single-cell RNA-Seq
  based on a multinomial model}.
\newblock \emph{{Genome Biology}}, 20\penalty0 (1):\penalty0 1--16, 2019.

\bibitem[Ullmann et~al.(2022)Ullmann, Hennig, and
  Boulesteix]{ullmann2022validation}
Theresa Ullmann, Christian Hennig, and Anne-Laure Boulesteix.
\newblock Validation of cluster analysis results on validation data: A
  systematic framework.
\newblock \emph{Wiley Interdisciplinary Reviews: Data Mining and Knowledge
  Discovery}, 12\penalty0 (3):\penalty0 e1444, 2022.

\bibitem[Van~den Berge et~al.(2020)Van~den Berge, De~Bezieux, Street, Saelens,
  Cannoodt, Saeys, Dudoit, and Clement]{van2020trajectory}
Koen Van~den Berge, Hector~Roux De~Bezieux, Kelly Street, Wouter Saelens,
  Robrecht Cannoodt, Yvan Saeys, Sandrine Dudoit, and Lieven Clement.
\newblock Trajectory-based differential expression analysis for single-cell
  sequencing data.
\newblock \emph{Nature Communications}, 11\penalty0 (1):\penalty0 1--13, 2020.

\bibitem[Zhang et~al.(2019)Zhang, Kamath, and David]{zhang2019valid}
Jesse~M Zhang, Govinda~M Kamath, and N~Tse David.
\newblock {Valid post-clustering differential analysis for single-cell
  RNA-Seq}.
\newblock \emph{Cell Systems}, 9\penalty0 (4):\penalty0 383--392, 2019.

\end{thebibliography}

\appendix
\section{Implementation details for Figure~\ref{fig_intro}}
\label{appendix_naive_sampsplit}

To generate Figure~\ref{fig_intro}, we first generate a toy dataset $X \in \mathbb{Z}_{\geq 0}^{100 \times 2}$, where each element $X_{ij}$ is drawn independently from the $\mathrm{NB}(5, 5)$ distribution, which has mean $5$ and variance $10$ (see Section~\ref{subsec_nb} for parameterization details). In panels (c) and (d), the naive method that uses the data twice  and our proposed method are implemented as described in Section~\ref{section_simulation_nb}. Here, we provide details about the implementation of sample splitting used in Figure~\ref{fig_intro}. 

We begin by splitting the cells such that the first $n/2$ cells belong to the training set, and the remaining cells belong to the test set, where $n=100$. 

We first describe Figure~\ref{fig_intro}(c). For values of $k$ ranging from $1$ to $10$, we cluster the first $n/2$ rows of the matrix $\log(X+1)$ using k-means clustering with $K$ clusters. This yields estimated cluster assignments $\hat{c}_i$ only for cells $i = 1,\ldots,n/2$. We then apply $3$-nearest neighbors classification to label each cell in the test set with the majority-label from its three nearest neighbors in the training set (using Euclidean distance on the log-transformed data).  We then compute an estimated mean $\hat{\mu}_{ij}$ for each datapoint $X_{ij}$ as
\begin{equation}
\label{eq_sampsplit_hatmu}
\hat{\mu}_{ij} = \frac{1}{\sum_{i'=1}^{n/2} \bold{1}\{ \hat{c}_i' = \hat{c}_i\}} \sum_{i'=1}^{n/2} X_{i',j}  \bold{1}\{ \hat{c}_i' = \hat{c}_i\},
\end{equation}
which is the sample mean of the training set data points belonging to this cluster. Finally, we compute the within-cluster mean-squared error as
\begin{equation}
\label{seq_sampsplit_mse}
\frac{1}{n/2 \times 2} \sum_{i = n/2+1}^n \sum_{j=1}^{2} \left( \log(X_{ij}+1) - \log(\hat{\mu}_{ij}+1) \right)^2.
\end{equation}

Sample splitting is implemented in a similar way in Figure~\ref{fig_intro}(d). We run $k$-means clustering with $k=2$ on the logged training data. This yields cluster assignments $\hat{c}_i$ only for $i=1,\ldots,n/2$. We once again obtain cluster assignments $\hat{c}_i$ for $i=n/2+1,\ldots,n$ by applying $3$-nearest neighbors. Finally, for $j=1$ and for $j=2$, we fit a negative binomial generalized linear model where the response is $X_{ij}$ and the covariate is $\hat{c}_i$, for $i=n/2+1, \ldots, n$. 

In both cases, sample splitting fails because $\hat{c}_i$ for $i=n/2+1, \ldots,n$ is obtained using the data $X_{ij}$ for $i=n/2+1, \ldots,n$, via the $3$-nearest neighbors classification step. Thus, the test set is not truly ``held out" in computing $\hat{c}_{i}$ for $i = n/2+1,\ldots,n$.

To create panels (e) and (f) of Figure~\ref{fig_intro}, we apply Algorithms~\ref{alg_cao_intra} and \ref{alg_cs_intra} to the single realization of toy data shown in Figure~\ref{fig_intro}. In both algorithms, we estimate $k=5$ clusters by running $k$-means on the log-transformed data. For Algorithm~\ref{alg_cao_intra}, we use a support vector machine (SVM) with a linear kernel as the classifier. 
\section{Proofs for Section~\ref{section_gcs}}
\label{appendix_proofs}

\subsection{Proof of Theorem~\ref{theorem_nbthin}}
\label{appendix_mainproof}

This result follows from applying Theorem 3 from \cite{neufeld2023data} to each individual element $X_{ij}$ in the matrix $X$, in the specific case where the $X_{ij}$ are independent negative binomial random variables. We note that \cite{neufeld2023data} use a different parameterization of the negative binomial: for $\by \sim \NB(r,p)$, they have $\E[\by] =  r \frac{1-p}{p}$ and $\Var(\by) =   r \frac{1-p}{p^2}$. This parameterization of the negative binomial is convolution-closed in the parameter $r$ if $p$ is held fixed, and corresponds to our parameterization if $\mu =r \frac{1-p}{p}$ and $b = r$. 

\subsection{Proof of Theorem~\ref{theorem_nb_binom_thin}}
\label{appendix_infproof}

We note that, when $b_j = \infty$, for $m = 1,\ldots,M$, $\bx_{ij}^{(m)} \mid \bx_{ij} = X_{ij} \sim \mathrm{Binomial}(X_{ij}, \epsilon_m)$ and $X^{(-m)} \mid \bx_{ij} = X_{ij} \sim \mathrm{Binomial}(X_{ij}, 1-\epsilon_m)$. Armed with these two facts, the first statement of Theorem~\ref{theorem_nb_binom_thin} is proved in \cite{harremoes2010thinning} and \cite{leiner2021data}, and the second statement of Theorem~\ref{theorem_nb_binom_thin} is proved in \cite{neufeld2022inference}. 

\subsection{Proof of Theorem~\ref{theorem_generalcase}}
\label{appendix_generalproof}

While parts of this theorem were proved in \cite{neufeld2022inference}, we prove this theorem in full directly here so that the notation matches that of this paper. 

The first statement of Theorem~\ref{theorem_generalcase} follows directly from the law of total expectation. Regardless of the value of $b'_j$, 
\begin{align*}
\E\left[\bxm_{ij}\right] &= \E\left[\E\left[\bxm_{ij} \mid \bx=X_{ij}\right]\right] = \E\left[\epsilon_m \bx_{ij}\right] = \epsilon_m \mu_{ij}, \\
\E\left[\bxmm_{ij}\right] &= \E\left[\E\left[\bxmm_{ij} \mid \bx=X_{ij}\right]\right] = \E\left[(1-\epsilon_m) \bx_{ij}\right] = (1-\epsilon_m) \mu_{ij}.
\end{align*}
This is because the parameters $b_j'$ do not affect the expected values of the \\ $\mathrm{DirichletMultinomial}\left(X_{ij}, \epsilon_1 b_j',\ldots, \epsilon_M b_j' \right)$ distribution, only the variance. \\
\\
The second statement of Theorem~\ref{theorem_generalcase} uses the law of total variance. We start by deriving the marginal variance of $\Var\left(\bxm_{ij}\right)$, as follows:
\begin{align*}
	\Var\left(\bxm_{ij}\right) &= \E\left[\Var\left(\bxm_{ij} \mid \bx_{ij} = X_{ij}\right)\right] + \Var\left( \E\left[ \bxm_{ij} \mid \bx_{ij} = X_{ij} \right]\right).
	\end{align*}
As $\bxm_{ij} \mid \bx_{ij} = X_{ij} \sim \bb\left(X_{ij}, \epsilon_m b_j', (1-\epsilon_m) b_j' \right)$, we plug in the (known) mean and variance of the beta-binomial distribution. 
\begin{align*}
\Var\left(\bxm_{ij}\right) &= \E\left[ \frac{\bx_{ij} \epsilon_m (1-\epsilon_m)  (b_j'+\bx)}{(b_j'+1)}\right] + \Var\left( \epsilon_m \bx_{ij} \right) \\
&= \frac{ \epsilon_m (1-\epsilon_m)  b_j' }{(b_j'+1)} \E\left[ \bx \right] + \frac{ \epsilon_m (1-\epsilon_m)}{(b_j'+1)} \E\left[ \bx^2 \right] + \epsilon_m^2 \Var\left( \bx_{ij} \right).
\end{align*}
We next plug in the known mean and variance of $\bx_{ij} \sim \NB(\mu_{ij},b_j)$.
\begin{align*}
\Var\left(\bxm_{ij}\right) &= \frac{ \epsilon_m (1-\epsilon_m)  b_j' \mu_{ij} }{(b'+1)} + \frac{ \epsilon_m (1-\epsilon_m)}{(b_j'+1)} \left(\mu_{ij} + \frac{\mu_{ij}^2}{b_j} + \mu_{ij}^2 \right) + \epsilon_m^2 \left(\mu_{ij} + \frac{\mu_{ij}^2}{b_j} \right) \\
&= \epsilon_{m} \mu_{ij} + \frac{\epsilon_{m}^2 \mu_{ij}^2}{b_j} 
+ \frac{ \epsilon_{m} (1-\epsilon_{m}) \mu^2}{(b_j'+1)} \left(\frac{1}{b_j} + 1 \right).
\end{align*}
To compare the magnitude of this variance to 
$\epsilon_m \Var(\bx)$, we add and subtract $\epsilon_m \frac{\mu_{ij}^2}{b_j}$.
\begin{align*}
\Var\left(\bxm_{ij}\right) &= \left[\epsilon_m \mu_{ij} + \epsilon_m \frac{\mu_{ij}^2}{b_j}\right] - \left[ \epsilon_m \frac{\mu_{ij}^2}{b_j} - \epsilon_m^2 \frac{\mu_{ij}^2}{b_j}\right] + \frac{ \epsilon_m (1-\epsilon_m) \mu_{ij}^2}{(b_j'+1)} \left(\frac{1}{b_j} + 1 \right) \\
&= \epsilon_m \Var(\bx) - 
\left[\epsilon_m (1-\epsilon_m) \frac{\mu_{ij}^2}{b_j} \right] + 
\frac{ \epsilon_m (1-\epsilon_m) \mu_{ij}^2}{(b_j'+1)} \left(\frac{1}{b_j} + 1\right) \\
&= \epsilon_m \Var(\bx) + \epsilon_m (1-\epsilon_m) \frac{\mu_{ij}^2}{b_j}\left( 
\frac{b_j+1}{b_j'+1} - 1\right),
\end{align*}
as claimed in Theorem~\ref{theorem_generalcase}. We omit the derivation of $\Var\left( \bxmm_{ij}\right)$, as it is identical to the derivation above after noting that $\Var(\bxmm_{ij} \mid \bold{X}_{ij} = X_{ij}) = \Var(\bx_{ij} - \bxm_{ij} \mid \bold{X}_{ij} = X_{ij}) = \Var(\bxm_{ij} \mid \bold{X}_{ij} = X_{ij})$.  \\
\\
Finally, for the third statement of Theorem~\ref{theorem_generalcase}, we use the fact that
$$
2 \times \mathrm{Cov}(\bxmm_{ij}, \bxm_{ij}) = \Var(\bold{X}_{ij})-\Var(\bxmm_{ij})-\Var\left(\bxm_{ij}\right).
$$
We then plug in the known values of these variances, and 
simplify, as follows:
\begin{align*}
2 \times \mathrm{Cov}(\bxmm_{ij}, \bxm_{ij}) &= \Var(\bold{X}_{ij}) - \epsilon_m \Var(\bold{X}_{ij}) - \epsilon_m (1-\epsilon_m) \frac{\mu_{ij}^2}{b_j} \left( 
\frac{b_j+1}{b_j'+1} - 1\right)\\
& \ \ \ \ \ - (1-\epsilon_m) \Var(\bold{X}_{ij}) - \epsilon_m (1-\epsilon_m) \frac{\mu_{ij}^2}{b_j} \left( 
\frac{b_j+1}{b_j'+1} - 1\right)
\\
&=  - 2 \epsilon_m (1-\epsilon_m) \frac{\mu_{ij}^2}{b_j} \left( 
\frac{b_j+1}{b_j'+1} - 1\right).
\end{align*}
Thus, $
\mathrm{Cov}(\bxmm_{ij}, \bxm_{ij}) = \epsilon_m (1-\epsilon_m) \frac{\mu_{ij}^2}{b_j} \left( 
1- \frac{b_j+1}{b_j'+1} \right)$.

\subsection{Proof of Theorem~\ref{theorem_fisher}}
\label{appendix_fisherproof}

We first derive the Fisher information contained in $X_{ij}$ for the parameter $\mu_{ij}$:
\begin{align*}
I_{\mu_{ij}}\left(\bold{X}_{ij}\right)	&= -\E \left[ \frac{d^2}{d \mu_{ij}^2} \log \left( f\left( \bold{X}_{ij} \mid \mu_{ij}, b_j \right) \right) \right] \\
&= -\E \left[ \frac{d^2}{d \mu_{ij}^2} \left( \log \left(\frac{\Gamma(\bold{X}_{ij}+b_j)}{\Gamma(b_j)\bold{X}_{ij}!} \right) + \bold{X}_{ij}  \log \left(\frac{\mu_{ij}}{\mu_{ij}+b_j}\right) + b_j \log \left(\frac{b_j}{\mu_{ij}+b_j}\right) \right)  \right] \\
&= -\E \left[ \bold{X}_{ij}  \left( \frac{-1}{\mu_{ij}^2} + \frac{1}{(\mu_{ij}+b_j)^2} \right) + b_j \left( \frac{1}{(\mu_{ij}+b_j)^2} \right) \right] \\
&=  \frac{1}{\mu_{ij}} - \frac{\mu_{ij}+b_j}{(\mu_{ij}+b_j)^2} 
=  \frac{ b_j }{\mu_{ij}(\mu_{ij}+b_j)}.
\end{align*}
Now we derive the Fisher information contained in $\bxm_{ij}$ for the parameter $\mu_{ij}$:
\begin{align*}
I_{\mu_{ij}}\left(\bxm_{ij} \right)	&= -\E \left[ \frac{d^2}{d \mu_{ij}^2} \log \left( f\left( \bxm_{ij} \mid \eps \mu_{ij}, \eps b_j \right) \right) \right] \\
&= -\E \left[ \frac{d^2}{d \mu_{ij}^2} \left( \log \left(\frac{\Gamma(\bold{X}_{ij}^{(m)}+\epsilon_m b_j)}{\Gamma(\epsilon_m b_j)(\bold{X}_{ij}^{(m)})!} \right) + \bold{X}^{(m)}_{ij}  \log \left(\frac{\epsilon_m  \mu_{ij}}{\epsilon_m \mu_{ij}+\epsilon_m b_j}\right) + \epsilon_m  b_j \log \left(\frac{\epsilon_m  b_j}{\epsilon_m \mu_{ij}+\epsilon_m b_j}\right) \right)  \right] \\
&= -\E \left[ \bold{X}_{ij}^{(m)}  \left( \frac{-1}{\mu_{ij}^2} + \frac{1}{(\mu_{ij}+b_j)^2} \right) + \epsilon_m b_j \left( \frac{1}{(\mu_{ij}+b_j)^2} \right) \right] \\
&=  \frac{\epsilon_m}{\mu_{ij}} - \frac{\epsilon_m \mu_{ij}+\epsilon_m b_j}{(\mu_{ij}+b_j)^2} 
=  \epsilon_m I_{\mu_{ij}}\left(\bold{X}_{ij}\right).
\end{align*}

Theorem~\ref{theorem_nbthin} guarantees independence between $\bxm_{ij} $ and $\bxmm_{ij}$, and therefore  $I_{\mu_{ij}}\left(\bxmm_{ij} \right)	+ I_{\mu_{ij}}\left(\bxm_{ij} \right) = I_{\mu_{ij}}\left(\bold{X}_{ij} \right)$. It follows that $I_{\mu_{ij}}\left(\bxmm_{ij} \right) =  (1-\epsilon_m) I_{\mu_{ij}}\left(\bold{X}_{ij}\right)$. 

\section{Implementation details for Section~\ref{section_simulation_nb}}
\label{appendix_sct}

We use the \texttt{R} package \texttt{sctransform} in the following manner to estimate gene-specific overdispersion parameters in the simulations for Section~\ref{section_simulation_nb}. 

Briefly, for each gene $j=1,\ldots,p$, \texttt{sctransform} begins by fitting a negative binomial GLM with $\bold{X}_j$ as the response, and the logged total number of unique molecular identifiers (UMIs) for the $n$ cells as the covariate. (In our simulations, the total number of UMIs for a cell is the row sum for that cell.) This yields a maximum likelihood estimate $\hat{b}_j^\mathrm{MLE}$ for each gene $j=1,\ldots,p$. These maximum likelihood estimates are known to be quite noisy for sparse negative binomial data.  Furthermore, the ``null model" that only includes the total number of UMIs as a covariate may be the correct model for 
the majority of the genes, but will be incorrect for any genes that exhibit true differential expression across unknown latent variables. Thus, as a second step,  \texttt{sctransform} fits a smooth kernel regression to estimate a relationship between the average expression of each gene and the gene-specific overdispersion. These smoothed estimates are used as the gene-specific overdispersion parameters. 

We run the \texttt{vst()} function from the \texttt{sctransform} package in \texttt{R} with its default settings in Section~\ref{section_simulation_nb}. We note that we simulated data in which the two main assumptions of \texttt{sctransform} are met: most genes are not differentially expressed, and there is a smooth relationship between the average expression of a gene and its parameter $b_j$. We use the same strategy in Section~\ref{section_realData_nb}, although in the real data setting we do not know for sure that the modeling assumptions of \texttt{sctransform} are met. 

\section{Implementation details for Section~\ref{section_realData_nb}}
\label{appendix_cao}

In Section~\ref{section_realData_nb}, we analyze a publicly-available dataset that is associated with \cite{cao2020human}, which can be downloaded from \url{https://descartes.brotmanbaty.org/}. For the main cell type analysis, we used all cells from this dataset that were collected from the kidney. For the cell subtype analysis, we used all of the kidney cells that were labeled as metanephric cells in the original analysis by \cite{cao2020human}. For both analyses, we filtered to genes with non-zero counts in at least 10 cells. After this subsetting, the kidney dataset has dimension $178,603 \times 34,714$ and the metanephric dataset has dimension $90,876 \times 31,385$. 

To carry out Step 1 of Algorithm~\ref{alg_cao_intra} on our two datasets, we used the \texttt{Monocle3} package. To preprocess each dataset, we generated a 50-dimensional principal components embedding of each dataset and subsequently a 2-dimensional UMAP embedding using the default settings of the preprocessing functions in the \texttt{Monocle3} package. 
Next, we performed Leiden clustering using the the \texttt{Monocle3} clustering function. We chose a resolution parameter that gave a similar number of clusters to those obtained in the original paper. More specifically, we set the resolution parameter to $1 \times 10^{-6}$ for the full kidney dataset and $1 \times 10^{-5}$ for the metanephric cell subset. To carry out Step 2 of Algorithm~\ref{alg_cao_intra}, we split the 50-dimensional principal components embedding of the full count data from kidney and metanephric cells into five folds containing equal numbers of cells. We used the cluster labels, inferred as described above (on the full counts), as the ``true" cluster labels. For each of the five folds, we then trained a linear SVM model to predict the cluster assignment from 80\% of the embedded expression data. We generated a confusion matrix by comparing the ``true" labels to this trained model's predictions on the held-out subset. We note that this SVM is slightly different from that of \cite{cao2020human}, who trained their SVM using the whole transcriptome rather than the reduced-dimension embedding. 

To carry out the ``assume Poisson" version of Algorithm~\ref{alg_cs_intra} on our two datasets, we performed Poisson count splitting (Algorithm~\ref{alg_cs}, or, equivalently, Algorithm~\ref{alg_gcs} with $b'_j = \infty$) with $M = 2$ folds and $\epsilon_1 = \epsilon_2 = 0.5$  
on each dataset to obtain an $X^{(1)}$ and an $X^{(2)}$ for each of the datasets. We then followed the preprocessing and clustering procedures outlined above on each fold for each dataset, to obtain two clusterings for each dataset. To produce Figures~\ref{fig_realData_confusion}(b) and \ref{fig_realData_confusion}(f), we re-ordered the test set labels on the $y$-axis to make the confusion matrix as diagonal as possible (since the clusters are invariant to re-labeling).

Finally, to carry out the negative binomial version of Algorithm~\ref{alg_cs_intra} on our two datasets, we applied the \texttt{sctranstorm} package in \texttt{R} with its default parameters to estimate overdispersion values for each gene (see Appendix~\ref{appendix_sct}). We then performed negative binomial count splitting (Algorithm~\ref{alg_gcs}) with $M = 2$ folds and $\epsilon_1 = \epsilon_2 = 0.5$  on each dataset to obtain an $X^{(1)}$ and an $X^{(2)}$ for each of the datasets. We then proceeded as in the Poisson case.

\end{document}